\documentclass[pre,aps,twocolumn,showpacs,superscriptaddress,floatfix]{revtex4}
\providecommand{\diff}{\mathrm{d}}
\newcommand{\tdiff}[2]{{\mathrm{d}#1}/{\mathrm{d}#2}}
\providecommand{\vc}[1]{\mathbf{#1}}
\newcommand{\uvc}[1]{\mathbf{\hat{#1}}}
\newcommand{\uij}{\vc{u}_{ij}}
\newcommand{\rij}{\vc{r}_{ij}}
\newcommand{\uparl}{u_\parallel}

\newcommand{\uperp}{u_\perp}

\newcommand{\uparlij}{u_{\parallel,ij}}

\newcommand{\uperpij}{u_{\perp,ij}}

\newcommand{\phic}{\phi_\mathrm{c}}
\newcommand{\dm}{\mathcal{M}}
\usepackage{graphicx}
\usepackage[dvips,usenames]{color}
\usepackage[normalem]{ulem}

\begin{document}
\title{Jammed frictionless discs: connecting local and global response}

\author{Wouter G. Ellenbroek}
\affiliation{Instituut--Lorentz,
Universiteit Leiden, Postbus 9506, 2300 RA Leiden, The Netherlands}
\affiliation{Department of Physics and Astronomy,
University of Pennsylvania, Philadelphia, PA 19104-6396, USA.}
\author{Martin van Hecke}
\affiliation{Kamerlingh Onnes Lab, Leiden University, PO box 9504,
2300 RA Leiden, The Netherlands.}
\author{Wim van Saarloos} \affiliation{Instituut--Lorentz,
Universiteit Leiden, Postbus 9506, 2300 RA Leiden, The
Netherlands}

\date{\today}

\begin{abstract}
By calculating the linear response of packings of soft
frictionless discs to quasistatic external perturbations, we
investigate the critical scaling behavior of their elastic
properties and non-affine deformations as a function of the
distance to jamming. Averaged over an ensemble of similar
packings, these systems are well described by elasticity, while in
single packings we determine a diverging length scale $\ell^*$ up
to which the response of the system is dominated by the local
packing disorder. This length scale, which we observe directly,
diverges as $1/\Delta z$, where $\Delta z$ is the difference
between contact number and its isostatic value, and appears to
scale identically to the length scale which had been
introduced earlier in the interpretation of the spectrum of
vibrational modes. It governs the crossover from isostatic
behavior at the small scale to continuum behavior at the large
scale; indeed we identify this length scale with the coarse
graining length needed to obtain a smooth stress field. We
characterize the non-affine displacements of the particles using
the \emph{displacement angle distribution}, a local measure for
the amount of relative sliding, and analyze the connection between
local relative displacements and the elastic moduli.
\end{abstract}
\maketitle

The jamming transition governs the onset of rigidity in disordered
media as diverse as foams, colloidal suspensions, granular media
and glasses ~\cite{jamnature,trappe}. While jamming in these
systems is controlled by a combination of density, shear stress
and temperature, most progress has been made for simple models of
frictionless soft spheres that interact through purely repulsive
contact forces, and that are at zero temperature and zero load
\cite{epitome,silbertPRL05,silbertPRE06,wyartEPL,wyartPRE,respprl}.
This constitutes possibly the simplest model for jamming.
Moreover, this is a suitable model for static foams or emulsions
\cite{bolton,durianbubble,weairebook}, which also represents a
simplified version of granular media, ignoring friction
\cite{maksePRL,ellakjam,kostyaPRE07} and non-spherical shapes
\cite{donev04,wouterse07,zorana09,bulbul09}.

From a theoretical point of view, this model is ideal for two
reasons. First, it exhibits a well defined jamming point, ``point J'', at confining pressure $p=0$, and in the limit of
large system sizes, the jamming transition occurs for a
well-defined density $\phi=\phi_c$, which has been identified with
the random close packing density~\cite{epitome}. At this
jamming point, the system is a disordered packing of frictionless
undeformed spheres, which is marginally stable and isostatic,
i.e.\ its contact number (average number of contacts per
particle) $z$ equals $z_\mathrm{iso}=2d$ in $d$ dimensions.
Second, in recent years it has been uncovered that the mechanical
and geometric properties of such jammed packings of soft spheres
close to point J exhibit a number of non-trivial power law
scalings as a function of $\Delta \phi:= \phi-\phic$. These
scalings illustrate the unique character of the jamming
transition
~\cite{epitome,silbertPRL05,silbertPRE06,wyartEPL,wyartPRE,respprl}.

Approaching point J from the jammed side, the most important
scaling relations are as follows. \emph{(1)} The excess contact
number $\Delta z:= z-z_{\rm iso}$ scales as $(\Delta \phi)^{1/2}$
\cite{durianbubble,maksePRL,epitome,wyartPRE}. \emph{(2)} The ratio
of shear ($G$) and bulk ($K$) elastic moduli vanishes at point J
as $G/K \sim \Delta z$ \cite{epitome}. \emph{(3)} The density of
vibrational modes exhibits a crossover from continuum like
behavior to a plateau at a characteristic frequency $\omega^*$,
which vanishes at the jamming point: $\omega^* \sim \Delta z$
\cite{silbertPRL05,wyartEPL,wyartPRE}. One can associate a
diverging length-scale $\ell^*$ with this crossover, which then
diverges at point J: $\ell^* \sim 1/\Delta z$ \cite{wyartEPL}.

Recently, we uncovered an additional non-trivial scaling near
point J when identifying the degree of non-affinity
\cite{respprl}. Decomposing, for linear deformations of jammed
systems, the relative displacement $\uij$ of neighboring particles
$i$ and $j$ in components parallel $(\uparl)$ and perpendicular
$(\uperp)$ to $\rij$, where $\rij$ connects the centers of
particles $i$ and $j$, we find that the ratio $\uperp/\uparl$
diverges near point J. More precisely, defining local displacement
angles $\alpha_{ij}$ via
$\tan\alpha_{ij}={u_{\perp,ij}}/{u_{\parallel,ij}}$, we find that
the displacement angle distribution $P(\alpha)$ peaks around
$\alpha=\pi/2$, with the peak height diverging near point J.
Hence, close to point J, the non-affinity is such that particles
predominantly \emph{slide past} each other.

The nature of the non-affinity of particle displacements ties in
with the question to what extent these disordered solids close to
the jamming transition can be described using continuum
elasticity. Recent work on granular materials has shown that this
is possible, using the proper definitions of stress and strain
tensor~\cite{goldcoarse}, and provided one probes the system at a
large enough length scale~\cite{goldPRL02,goldnature,goldPRL06}.
However, it has remained unclear if this length scale is to be
identified with the crossover length scale $\ell^*$ that was
introduced in the interpretation of the density of vibrational
states~\cite{wyartEPL}.  In addition, despite the large body of
numerical and experimental work on the
problem~\cite{vanel,serero01,gengPRL01,reydellet01,ottoPRE03,atmanbrunet,srdjanPRL06},
a systematic analysis of the relations between the bulk elastic
moduli, the stress fields in response to local perturbations, and
the distance to the jamming transition appears to be lacking.

In this paper, we describe the applicability of elasticity theory for jammed
packings, and elaborate on our earlier work on the displacement angle
distributions~\cite{respprl}.
Although all our numerical studies are restricted to
frictionless spheres in two dimensions, we expect essentially all concepts and
conclusions to carry over to three dimensions as well.

In section I we detail our linear response methodology, and
introduce our notation. In section II we focus on the response of
jammed systems to local and global forcing, and show that this
response, suitably coarse grained and averaged, can be described
by linear elasticity. Earlier direct numerical simulations
\cite{epitome,maksePRE} have established the scalings of  the
shear modulus $G$ and bulk modulus $K$ with distance to
jamming, and in particular have shown that their ratio $G/K$ vanishes at
point J. Our linear response calculations reproduce these
findings.

In section III we address the issue of the length scale beyond
which elasticity theory can be applied to the system. First, we
determine this length directly from the response to a local
perturbation, as the scale up to which the response is dominated
by spatial fluctuations, and identify it with $\ell^*\sim1/\Delta
z$~\cite{wyartEPL}. In addition, we show that the coarse graining
length, needed to obtain a smooth response in a globally deformed
system, is proportional to the same length $\ell^*$.

In section IV we characterize the non-affine nature of the
response to bulk deformations, and elaborate on the scaling of the
displacement angle distribution $P(\alpha)$. We discuss the
connection between this scaling and the form of the expression for
the changes in elastic energy in linear order: the opposite sign
of the contribution from the parallel $(\uparl)$ and
perpendicular $(\uperp)$ components of the local displacements
naturally leads to a delicate balancing act for the case of an
overall shear deformation, while compression leads to a more
convoluted picture~\cite{wouterEPL09}.

\section{Linear Response}
All numerical results presented in
this paper concern the quasistatic linear response of a certain
starting configuration to a small perturbation. The
approach is therefore explicitly a two-step one. First, we prepare a
two-dimensional packing of frictionless polydisperse discs at a
given density or confining pressure, using a Molecular Dynamics
simulation.  The resulting packings are then analyzed by studying their
response to small perturbations. Response studies have been
done previously by using MD also for the perturbed system~\cite{maksePRE},
in which case the full dynamics are taken into
account. Another method that has been used is the quasistatic
approach of minimizing the potential energy, while ignoring
inertia~\cite{epitome}. We here take the even simpler approach
of explicitly focussing on the linear response equations. This
requires solving just a single matrix equation for each numerical
experiment, which makes it numerically cheap.

In the following subsections we describe the procedure to
generate our packings~\cite{ellakwave,ellakjam}, and recapitulate
the derivation of the linear response
equation~\cite{leonforte,wouterpowders,respprl} from an expansion
of the elastic energy of the system~\cite{tanguy,wyartPRE}, both
for completeness and to introduce the necessary notation.

\subsection{Packing Generation}
\label{packgen} We prepare our packings using a molecular dynamics
simulation with round discs in two dimensions. The discs interact
via the three-dimensional Hertzian potential,
\begin{equation}
\label{hertzpot} V_{ij}=\left\{
\begin{array}{l@{\qquad}r}
\epsilon_{ij}\left(1-\frac{r_{ij}}{d_{ij}}\right)^{5/2}
&r_{ij}\leq d_{ij}\\
0&r_{ij}\geq d_{ij}
\end{array}
\right.~,
\end{equation}
where $i,j$ label the particles, $d_{ij}$ is the sum of the
particle radii $R_i+R_j$, and $r_{ij}$ is the distance between the
particle centers. The energy scale $\epsilon_{ij}$ depends on the
radii and effective Young moduli of the particles, see
Appendix~\ref{app.Hertz}. The quantity between brackets is the
dimensionless overlap, $\delta_{ij}=1-r_{ij}/d_{ij}$.

Using the 3D Hertzian potential makes the system a quasi-2D
packing of discs with round edges.  We use zero gravity to have a
homogeneous pressure, and the radii of the discs are drawn from a
flat distribution between $0.4<R<0.6$, thus creating a
polydispersity of $\pm20\%$ around the average particle diameter
$d=1$ (our unit of length) to avoid crystallization. The
simulation starts from a loose granular gas in a square box with
periodic boundaries, which is compressed to a target pressure $p$
by changing the radii of the particles while they are moving
around. The radii are multiplied by a common scale factor $r_s$,
which evolves in time via the damped equation
$\ddot{r}_s=-4\omega_0\dot{r}_s-\omega_0^2 [p(t,r_s)/p-1]r_s$,
where $\omega_0 \sim 6\cdot10^{-2}$, $p(t,r_s)$ is the
instantaneous value of the pressure and $p$ the target pressure.
This ensures a very gentle equilibration of the packings.  Energy
is dissipated through inelastic collisions and a drag force which
slows down the particles. The simulation stops when the
accelerations of all grains have dropped below a threshold which
is $10^{-6}\langle f\rangle$ in our reduced units. This way we
generate packings in mechanical equilibrium for pressures ranging
from $p=10^{-6}$ to $p=3\cdot 10^{-2}$, in units of the modified
Young modulus of the individual grains~\cite{contact}. This
corresponds to a range of contact numbers from $z=4.05$ to
$z=5.87$. For one particular type of calculation we use packings
which are only periodic in the $x$-direction, and have hard walls
on top and bottom.  These are generated by compressing the system
using a hard piston, as described in Ref.~\cite{ellakwave}.

\begin{figure}
\includegraphics[width=6cm]{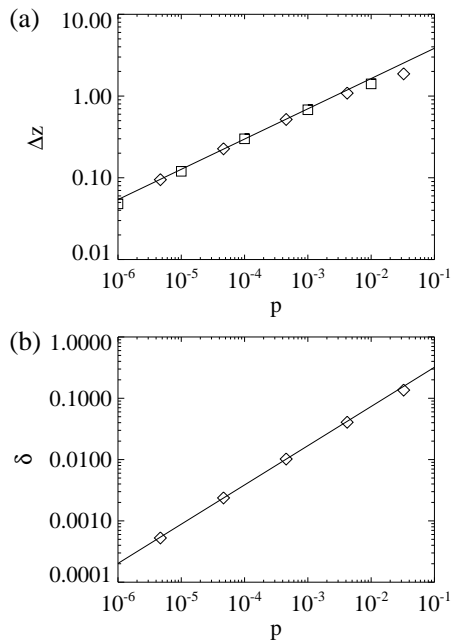}
\caption{{The distance to the jamming transition is set in our
packing preparation by the pressure $p$. Other parametrizations
are (a) the distance to isostaticity $\Delta
z=z-z_\mathrm{iso}\approx9\cdot p^{0.37}$ and (b) the typical
dimensionless interparticle overlap $\delta\approx1.4\cdot
p^{0.64}$. The solid lines represent these empirical power law
fits. Diamonds represent periodic packings, and squares are for
packings with hard walls on top and bottom.}}
\label{fig:packparam}
\end{figure}
Because there is no gravity in our simulations, at the end of the
simulation there will usually be particles without neighbors, or
due to numerical precision effects, particles with fewer neighbors
than is needed to make them rigidly connected to the rest of the
packing. These \emph{rattlers} or \emph{floaters} do not
contribute to the rigid backbone of the packing. They are removed
from the system when determining the contact number $z$ or
calculating its linear response to small perturbations.

To express the distance to the jamming point, we will use the
pressure $p$ and typical relative overlap $\delta$
interchangeably; the pressure (rather than the density) is most
conveniently set in the numerical procedure to generate the
packings, while the overlap enters in the scaling relations at the
particle scale that will be discussed in section~\ref{sectionenergyminimization}.
We do not use $\Delta\phi=\phi-\phi_c$, because in finite systems
$\phi_c$ is a number that would have to be determined for each
packing separately. However, close to point J we have determined
that $\Delta\phi\sim\delta$.
The conversion between the different parametrizations is given
in Fig.~\ref{fig:packparam}.

\subsection{Linear response equation}
\label{subsec.linresp}
We calculate the response of a packing to a mechanical perturbation, which
can be either an external force or a deformation of the periodic box. The
response in general involves displacements of all particles in the
packing, which we analyze in the linear regime.

The total potential energy of the system is a sum over all pairs of
interacting particles,
\begin{equation}
\label{totalE}
E=\sum_{\langle i,j\rangle}V_{ij}(r_{ij}),
\end{equation}
where we assume we are dealing with a central potential, only depending on
the distance between the particles
$r_{ij}=|\vc{r}_{ij}|=|\vc{r}_j-\vc{r}_i|$.
The change in $r_{ij}$ due to
displacements $\vc{u}_i$ of the particles is
\begin{equation}
\label{Deltarij}
\Delta r_{ij}=\sqrt{(r_{ij}+\uparlij)^2+\uperpij^2}-r_{ij}~,
\end{equation}
where $\uparl$ is the relative displacement along the line
connecting the centers of the particles and $\uperp$ is the
relative displacement perpendicular to that. In the  linear
response regime the change in energy is expanded to second order
in $\uparl$ and $\uperp$ as
\begin{equation}
\label{DeltaE}
\Delta E=\frac12\sum_{\langle i,j\rangle}
k_{ij}\left(u_{\parallel,ij}^2-\frac{f_{ij}}{k_{ij}r_{ij}}u_{\perp,ij}^2\right)~.
\end{equation}
Here $f_{ij}=-\tdiff{V_{ij}}{r_{ij}}$ and $k_{ij}=\tdiff{^2
V_{ij}}{r_{ij}^2}$. For all (power law) interactions that are
reasonable models for foams or granular media (linear repulsion,
Hertzian repulsion), both the initial force $f$ and the stiffness
$k$ are nonnegative (see below). There are no terms linear in the
displacements because the starting configuration is in mechanical
equilibrium, which makes these terms sum to zero. The
$u_\parallel^2$-term represents the change in bond length. The
$u_\perp^2$-term is only there if there is a pre-stress or initial
force and
captures the lowering of the energy due to
transverse displacements of the particles~\cite{alexander}.

The interaction potential for the particles that make up the packings used
in this paper is a finite-range, purely repulsive potential of the form
$V\sim\delta_{ij}^\alpha$, where
\begin{equation}
\label{defdelta}
\delta_{ij}=1-\frac{r_{ij}}{R_i+R_j}
\end{equation}
is the dimensionless virtual overlap of the particles
and $R_i$ and $R_j$ are the radii of particles $i$ and $j$.
The exponent
$\alpha$ has the value $\alpha=5/2$ in our packings, representing the
three-dimensional Hertzian interaction. For any potential of
power law form, the
prefactor of the second term in equation~(\ref{DeltaE}) can be written as
$\delta_{ij}/(\alpha-1)$. The closer to point J, the smaller these
dimensionless overlaps, so this prefactor will be small close to point J.
However, this does not allow us to ignore this term, because, as we will
see, the typical $u_\perp$ are going to be much larger than the typical
$u_\parallel$, in the limit of approaching the jamming transition: the
two terms in Eq.~(\ref{DeltaE}) become, in some cases, of the same order.

Writing the change in energy in the independent variables of the
problem, the displacements $\vc{u}_i$ of the particles, defines the
dynamical matrix $\dm$:
\begin{equation}
\label{dynmatdef} \Delta
E=\frac12\dm_{ij,\alpha\beta}~u_{i,\alpha}~u_{j,\beta}~,
\end{equation}
where $\dm$ is a $dN\times dN$ matrix ($d$ being the spatial
dimension and $N$ the number of particles). The indices
$\alpha,\beta$ label the coordinate axes and we use the summation
convention. The dynamical matrix contains all information on the
elastic properties of the system. It can be diagonalized to study
the vibrational properties~\cite{epitome,tanguy,ellakjam}, but it
can also be used to study the response of the system to an
external force (defined on each
particle)~\cite{leonforte,wouterpowders,respprl}:
\begin{equation}
\label{responseeq}
\dm_{ij,\alpha\beta}~u_{j,\beta}=f^\mathrm{ext}_{i,\alpha}~.
\end{equation}
The dynamical matrix $\dm$ is very sparse for large systems, because the
only nonzero elements are those for which $i$ and $j$ are in contact
and those for which $i=j$. Therefore we can compute the response
efficiently by using the Conjugate Gradient Method~\cite{templates}. This is
essentially an iterative procedure that minimizes
$||\dm_{ij,\alpha\beta}u_{j,\beta}-f^\mathrm{ext}_{i,\alpha}||$. O'Hern~\emph{et al.}~\cite{epitome} also studied the quasistatic response of granular systems to
a global shear or compression~\cite{epitome} using a Conjugate Gradient
Method, but in their case the quantity to be minimized was the potential
energy.  The difference is therefore that
we use Conjugate Gradient only as an efficient method
to study small deviations from a stable starting
configuration in linear reponse,
while O'Hern \emph{et al.}\ used it for various situations where
they needed to find the nearest potential energy minimum.

\subsection{Forces and stresses}
Solving the linear response equation~(\ref{responseeq}) yields the
displacements $\vc{u}_i$ of all particles. From this we calculate the
local relative displacements $u_{\perp,ij}$ and $u_{\parallel,ij}$, and the
change in contact force $\Delta f_{ij}$ using
\begin{eqnarray}
\label{uparlcoord}
u_{\parallel,ij}&=&\cos\phi_{ij}u_{ij,x}+\sin\phi_{ij}u_{ij,y}~,\\
\label{uperpcoord}
u_{\perp,ij}&=&\cos\phi_{ij}u_{ij,y}-\sin\phi_{ij}u_{ij,x}~,\\
\label{dfcoord}
\Delta f_{ij}&=&-ku_{\parallel,ij}~,
\end{eqnarray}
where $\phi_{ij}$ is the angle which the bond vector $\vc{r}_{ij}$ makes
with the $x$-axis.

To get from discrete forces to a continuum stress field, we apply the local
coarse graining procedure developed by Goldhirsch and
Goldenberg~\cite{goldcoarse}:
\begin{equation}
\label{stresscoord}
\Delta\sigma_{\alpha\beta}(\vc{r})=\sum_{\langle
i,j\rangle}\Delta f_{ij,\alpha}r_{ij,\beta}
\int_0^1\diff s\,\Phi\left(\vc{r}-\vc{r}_j+s\vc{r}_{ij}\right)~.
\end{equation}
It has been shown that this procedure gives local stress fields that do not
depend strongly on the coarse graining length chosen --- consistent results
were obtained for length scales down to a single grain
diameter~\cite{goldPRL06}. In this paper, we use a Gaussian coarse graining
function $\Phi$ with a width $\xi_\mathrm{CG}$ equal to the average particle
diameter, $\xi_\mathrm{CG}=1$ in our units, i.e.
$$
\Phi(\vc{r})=\frac{1}{2\pi}e^{-|\vc{r}|^2/2}.
$$
If, instead, one uses $\Phi=1/V$, Eq.~(\ref{stresscoord}) reduces to the well-known expression
for the average of the contact stress tensor over the entire system.

The precise form of Eq.~(\ref{stresscoord}) is not crucial for the work presented in this
paper, because we will only use this coarse graining procedure for the
stresses, not for the strains. It is the requirement of emerging linear
elasticity from coarse graining both stress and strain that led to this
particular form in Ref.~\cite{goldcoarse}.
It should also be noted
that, in principle, the expression for $\Delta \sigma$ contains both terms
$\sim \Delta f~r$ and $\sim f ~\Delta r$ --- we ignore the latter since these
terms are negligibly small close to the jamming transition.

\section{Macroscopics and continuum elasticity}
\label{sec.macro} There are three questions that
will be discussed here
concerning the macroscopic aspects of the linear
response of granular packings. First of all, under what
conditions can the system's response to external forcing be
described using continuum elasticity? Secondly, do we recover,
in the linear regime, the scaling of the elastic moduli with
contact number that was obtained in direct numerical
simulations~\cite{epitome}? Finally, is there a difference
in the values of the elastic moduli as calculated from the
response to global forcing, using bulk compression or shear, and
local forcing, applying an external force on a single particle?

The work by Goldenberg and
Goldhirsch~\cite{goldcoarse,goldPRL02,goldnature,goldPRL06} has
made clear that stress and strain tensors are well-defined down to
even less than the scale of a single grain, using the coarse
graining expressed by Eq.~(\ref{stresscoord}). For large enough
systems, they find elasticity to provide a good description of the
response of granular media. In this section, we analyze the
response of granular packings \emph{as a function of distance to
the jamming transition}, and confirm that the linear response of
granular media, when averaged over an ensemble of similarly
prepared packings, is well described by continuum elasticity. To
do so, we study the response to both global and local
perturbations, as described in sections \ref{sec.bulkdef} and
\ref{sec.pointresp}, respectively. A fit of the ensemble averaged
stresses and displacements provides the elastic moduli of the
packings, which we find to scale consistently with earlier results
\cite{epitome}.

\subsection{Bulk deformations}
\label{sec.bulkdef}
\begin{figure*}
\includegraphics[width=\textwidth]{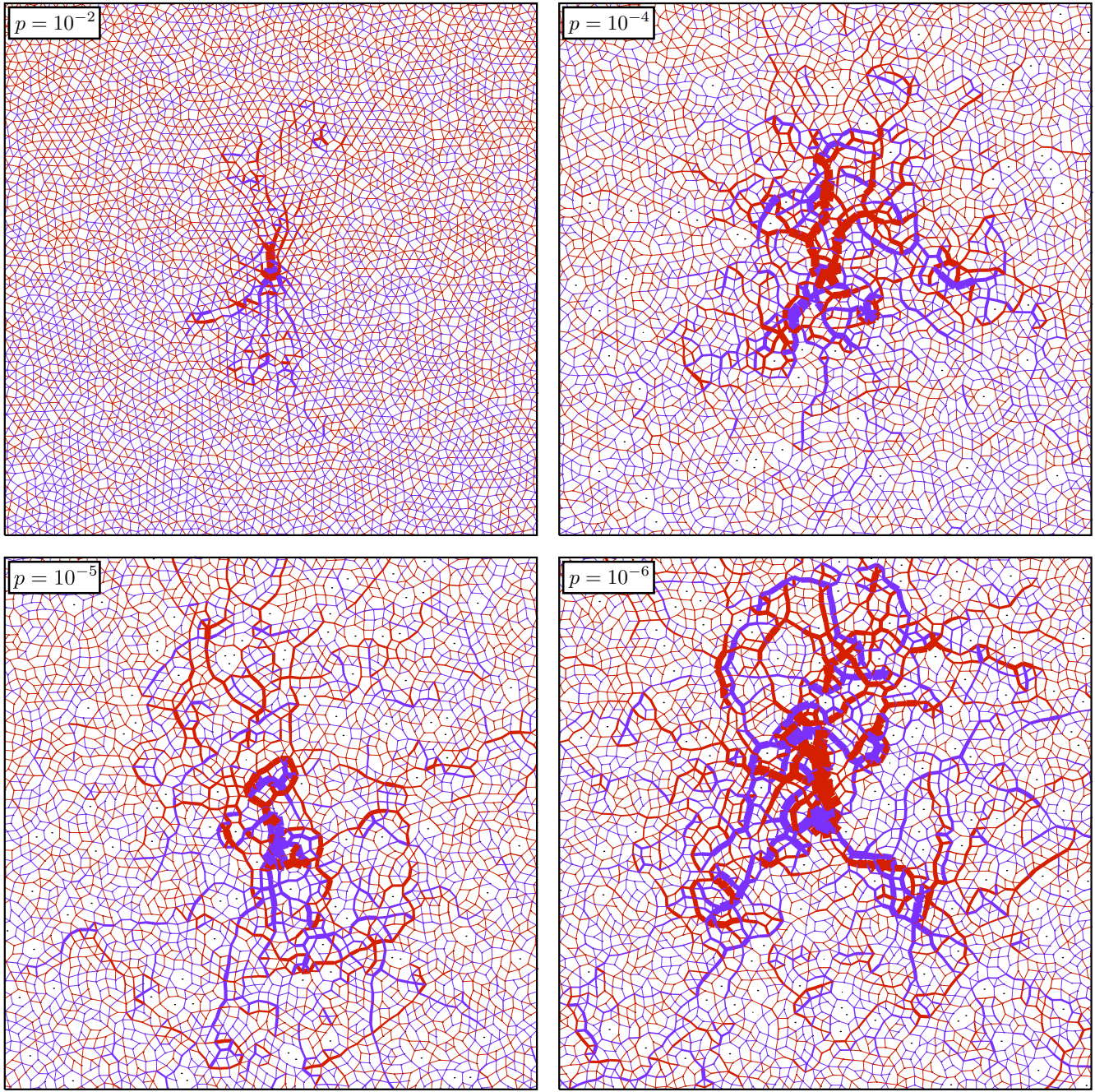}
\caption{(color) Force response networks for a point loading with
pressure as indicated. Blue (red) lines indicate an increase
(decrease) of the contact force, the thickness indicating the
amount (the thickness of the black border around each panel corresponds
to 1/15 of the loading force on the center particle).
The particles themselves are not drawn.  The figures show
only the part of the packing close to the forced grain, and
contain about 3500 of the 10\,000 particles.}
\label{fig:respnetworks}
\end{figure*}
The conventional way to extract elastic constants of a packing is
to apply a compression or shear deformation to the entire system.
In packings with periodic boundaries this is done by imposing a
relative displacement on all bonds that cross the boundary of the
periodic box,
following a procedure that is similar to the Lees-Edwards boundary conditions employed to simulate uniform shear flows~\cite{leesedwards}.
For example, to impose a globally uniform
compressional strain $\epsilon_{xx}=\epsilon_{yy}=\epsilon$, all
terms in the energy expansion of equation~(\ref{DeltaE}) that
represent a bond that crosses the periodic $x$-boundary are
changed such that each occurrence of $\vc{u}_j-\vc{u}_i$ is
replaced by $\vc{u}_j-\vc{u}_i\pm\epsilon L_x\uvc{x}$, where the
$\pm$-sign is given by the sign of $r_{ij,x}$.  The $y$-boundary
is treated analogously. Shear deformations are applied in the
form of a pure shear, i.e., by having a displacement in the
$y$-direction imposed on all bonds that cross the $x$-boundary
\emph{and} a displacement in the $x$-direction on all bonds that
cross the $y$-boundary.

When we write the linear
response equation~(\ref{responseeq}) from the energy expression
for the deformed system, the terms proportional to $\epsilon$ end
up on the right hand side of the equation and act like an
effective $f^\mathrm{ext}$. The response of the system to this
shape or volume change of the periodic box is again calculated by
solving equation~(\ref{responseeq}) for this effective external
force. The elastic moduli then follow from the resulting change in
stress tensor according to equation~(\ref{stresscoord}) with the
trivial coarse graining function $\Phi=1/V$. The bulk modulus is
extracted from a uniform compressional strain
$\epsilon_{xx}=\epsilon_{yy}=\epsilon$ through
\begin{equation}
\label{bulkmod}
K=\frac{\sigma_{\alpha\alpha}}{2\epsilon_{\alpha\alpha}}
=\frac{\sigma_{xx}+\sigma_{yy}}{4\epsilon}~,
\end{equation}
and the shear modulus from a uniform shear strain
$\epsilon_{xy}=\epsilon$ through
\begin{equation}
\label{shearmod}
G=\frac{\sigma_{xy}}{2\epsilon_{xy}}=\frac{\sigma_{xy}}{2\epsilon}~.
\end{equation}
The results will presented and discussed in section~\ref{sec.macrodiscuss}, together
with those obtained from the local point response.

\subsection{Point response}
\label{sec.pointresp} Now let us determine the linear response of
packings of $N=10\,000$ particles to a force in the $y$-direction,
applied to a single particle. The packings are periodic in the
$x$-direction and have hard walls on top and bottom to carry the
load. The confining pressure $p$ used to create the packings
ranges from $p=10^{-6}$ to $p=10^{-2}$, corresponding to contact
numbers ranging from $z=4.05$ to $z=5.41$. We have 10 different
packings at each pressure, and use each one about 20 times by
applying a point force to different particles, all of which are
closer than 0.1 particle diameter to the horizontal line through
the center of the system. Examples of the resulting \emph{response
networks}, depicting the changes in the contact forces, are shown
in Fig.~\ref{fig:respnetworks}. These pictures immediately
illustrate to the eye, that as the jamming point is approached,
both the range and the magnitude of the fluctuations increase.
Quantifying this behavior is the goal of
section~\ref{sec.lengthscale} --- here we focus first on the
ensemble averaged response.

\begin{figure}
\includegraphics[width=8.6cm]{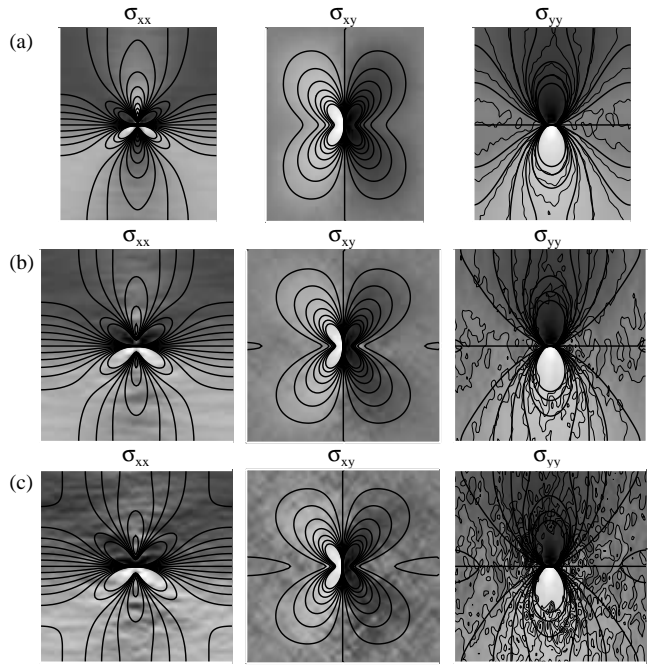}
\caption{Stress response fields for the linear response to a point force. The
greyscale images represent the ensemble averaged granular stress response
field; the thick contours denote the fitted continuum stress
field. For $\sigma_{yy}$ thin contours for the granular stress response
field are included as well. The components of the stress tensor
are plotted for (a) $p=10^{-2}$, (b) $p=10^{-4}$,  (c)
$p=10^{-6}$.}
\label{fig:stressfields}
\end{figure}

In order to allow comparison to continuum solutions, we first
calculate for each force response network the associated stress
response fields, by the coarse graining procedure in Eq.~(\ref{stresscoord}).
See Ref.~\cite{goldcoarse} for details. We then
calculate the ensemble average of these stress response fields,
and results of this procedure are shown in
Fig.~\ref{fig:stressfields}. The continuum solution that we are
comparing the granular point response to, is obtained from the
static Navier-Cauchy equation \cite{benny}, which is given by
\begin{equation}
\label{navcau}
G\Delta\vc{u}+K\nabla(\nabla\cdot\vc{u})=-\vc{f}^\mathrm{ext},
\end{equation}
in a two-dimensional formulation of elasticity theory. Here $K$
and $G$ denote bulk and shear modulus. This equation is the direct
equivalent of the linear response equation~(\ref{responseeq}),
because solving it yields displacements in terms of external
forces. We solve equation~(\ref{navcau}) with
$\vc{f}^\mathrm{ext}$ equal to a unit point force in the
$y$-direction; see Appendix~\ref{app.Elas} for details. The
resulting stress field $\sigma_{\alpha\beta}(\vc{r})$ only depends
on the ratio $K/G$ of elastic constants, since the overall scale
drops out when relating an imposed force to a resulting stress.
Therefore, there is only one free parameter when fitting the
stress field of the granular system, equation~(\ref{stresscoord}),
to the continuum expression. As described in
Appendix~\ref{app.Elas}, we use the the effective Poisson ratio
$\nu=(K-G)/(K+G)$. To determine both moduli we need a second
fitting parameter, using a fit to the displacement field. In
particular, we fit the the average $y$-displacement of all
particles in a strip at height $y$,
\begin{equation}
\label{granHy}
H_\mathrm{gran}(y):=\left\langle u_{i,y}\right\rangle_{y_i\approx y}~,
\end{equation}
to its continuum counterpart:
\begin{equation}
\label{exactHy}
H(y):=\frac{1}{L_x}\int_{-L_x/2}^{L_x/2}u_y(x,y)\text{d}x=\frac{L_y-2|y|}{4L_x(K+G)},
\end{equation}
which is obtained by taking
$\mathbf{f}^\mathrm{ext}=\delta(x)\delta(y)\uvc{y}$ and
integrating equation~(\ref{navcau}) over $x$, leaving a standard Green's
function problem with the boundary condition that $u_y$ vanishes at
the top and bottom wall, i.e., at $y=\pm L_y/2$.

Figure~\ref{fig:stressfields} displays the ensemble averaged stress response fields
$\Delta\sigma_{\alpha\beta}(\vc{r})$ (equation~(\ref{stresscoord})) and the
fitted continuum stress fields for $p=10^{-2}$, $p=10^{-4}$, and $p=10^{-6}$.
The grey scale and the contour values are chosen for each tensor component
separately, but for each component are the same for the different pressures.  The fit is very
good, especially considering the fact that a tensor field with three components
is fitted with only 1 parameter. For the $yy$-component we also include
contours for the numerical data, to give a more quantitative view of the fit.
For the $xx$- and $xy$-components the numerical data is too noisy for this to
be useful, especially at the lower pressures.

\begin{figure}
\includegraphics[width=7cm]{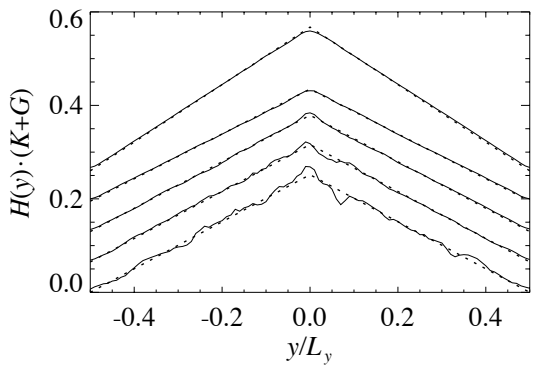}
\caption{The fitting procedure used to obtain $K+G$: The average vertical
displacement of particles at height $y$, equation~(\ref{granHy}), is fitted
to the triangle-shaped function $H(y)$, given by equation~(\ref{exactHy}).
The functions are rescaled by the fitted $K+G$ and vertically offset for
clarity (note that $H(y)=0$ for $y/L_y=\pm\frac12$).}
\label{fig:fitdisp}
\end{figure}
The fits of the average vertical displacements $H_\mathrm{gran}(y)$ are shown
in Fig.~\ref{fig:fitdisp}. Again a good fit is obtained with only one
parameter, with only a little bit of noise for the lowest pressure packings.
Combining the results of the two fitting procedures yields the elastic moduli
--- the results are presented in the next subsection.

\subsection{Results}
\label{sec.macrodiscuss}
\begin{figure}
\includegraphics[width=7cm]{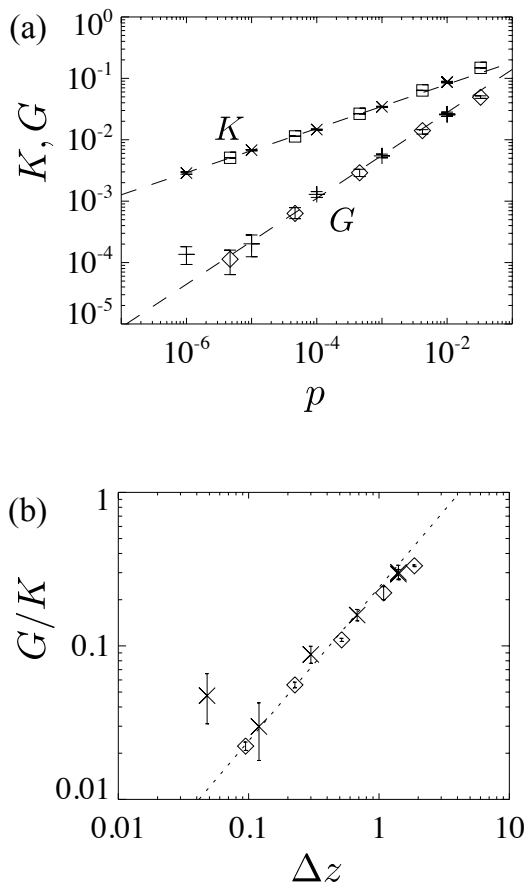}
\caption{(a) Bulk modulus $K$ and shear modulus $G$ as a function
of pressure. The squares ($K$) and diamonds ($G$) are obtained
using the bulk response described in section~\ref{sec.bulkdef}.
The crosses ($K$) and plus signs ($G$) follow from the fits of the
point response stresses and displacements discussed in section~\ref{sec.pointresp}.
The error bars on the bulk response data
indicate the intervals spanned by the median 50\% of the data; the
actual standard error of the mean is much smaller than the symbol
size. The error bars on the point response data are error
estimates from the fitting procedure. We attribute the fact that
the point response  result for $G$ at the smallest pressure
deviates from the scaling behavior to the fact that our fitting to
elastic continuum behavior is very insensitive to the value of $G$
in the limit $G \ll K$.
(b) The ratio of elastic moduli $G/K$ scales approximately as $\Delta z$.
}
\label{fig:moduli}
\end{figure}

The elastic moduli resulting from the methods explained in the
previous two subsections are collected in Fig.~\ref{fig:moduli}a.
The squares and diamonds represent the bulk ($K$) and shear ($G$)
modulus, respectively, calculated from the linear response to a
bulk compression or shear. These have been obtained via ensemble
averages over 100 packings of $N=1000$ particles, for pressures
ranging from $p=5\cdot10^{-6}$ to $p=3\cdot10^{-2}$ (in units of
the modified Young modulus of the grains). This pressure range
corresponds to a range in contact number from $z=4.10$ to
$z=5.87$. The cross and plus signs indicate the moduli resulting
from fitting the stress response to a point force to continuum
elasticity. As mentioned in the previous subsection, the packings
used for the point response calculations were prepared at
pressures ranging from $p=10^{-6}$ to $p=10^{-2}$, corresponding
to contact numbers from $z=4.05$ to $z=5.41$.

From Fig.~\ref{fig:moduli} one sees that the
data is well described~\footnote{The
point which deviates from the
scaling of $G$ is obtained for the lowest pressure from the point
response. We believe this large deviation to be due to the fitting
procedure: we fit the Poisson ratio $\nu$ from the stress field
$\Delta\sigma_{\alpha\beta}(\vc{r})$. As $p\to0$, we observe
$\nu\to1$, its maximum value in two dimensions. In this limit, the
fitting of the stress tensor becomes less accurate for numerical
reasons, which is made even worse by the fact that the data become
more noisy at low $p$. In the bulk response we see a similar
effect: the fluctuations around the average values are much larger
at lower pressure.} by scaling relations of the form
\begin{eqnarray}
K&\sim&p^{0.36\pm0.03}~,\label{Kscaling}\\
G&\sim&p^{0.70\pm0.08}~.\label{Gscaling}
\end{eqnarray}
These are consistent with the scalings $K\sim p^{1/3}$ and  $G\sim
p^{2/3}$ which on the basis of previous work \cite{epitome} are
expected for our 3D Hertizan contacts. The above scaling
behavior already shows that we can consistently describe the
response in terms of continuum elasticity.

Alternatively, we can
plot the ratio of the elastic moduli as function of $\Delta z$,
and find that $G/K$ scales as $\Delta z$ (Fig.~\ref{fig:moduli}b).
This result is expected to be independent of the force law --- in all
disordered packings that have been checked, the bulk modulus $K$
is proportional to the contact stiffness $k$, while the shear
modulus $G$ behaves proportional to $k\Delta
z$~\cite{epitome,maksePRE,wyartthesis}. The scaling of the shear
modulus has been described as
anomalous~\cite{maksePRL,maksePRE,epitome}, because earlier
\emph{effective medium theories} had predicted $K\sim G\sim
k$~\cite{walton87,maksePRL}. However, from the perspective of rigidity
percolation on a random network, for which all elastic moduli are proportional
to $\Delta z$, the compression modulus can be described as anomalous.  We
have recently explored this issue extensively in the context of harmonic
networks~\cite{wouterEPL09}.

To our knowledge, our calculations are the first to show that
the average large scale response to both a local force and a
global deformation or force is consistent with continuum
elasticity theory with the \emph{same} elastic moduli. We thus
conclude that the linear response to a local perturbation, \emph{averaged}
over an ensemble of packings, is well described by
linear elasticity. To what extent and on what length scale the
response of a \emph{single} packing corresponds to elastic behavior
will be the subject of section~\ref{sec.lengthscale}.

\begin{figure}
\includegraphics[width=8.5cm]{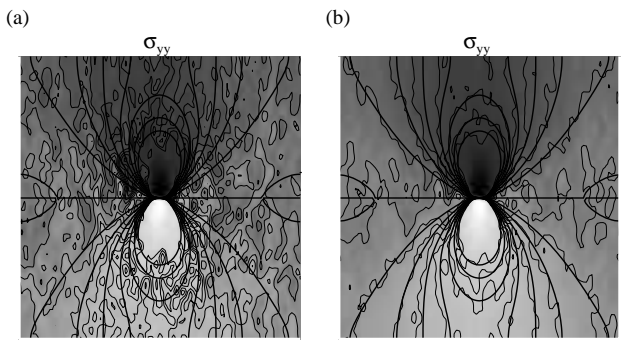}
\caption{Comparison of the stress response to a point force of a packing
close to the jamming transition ($p=10^{-6}$), (a) calculated with the energy
expansion as stated in Eq.~(\ref{DeltaE}), and (b) calculated without the
$\uperp^2$-term, which corresponds to ignoring the forces present in the
system before the point force was applied. Thick curves are the continuum
fit, thin curves the numerical contours, as in Fig.~\ref{fig:stressfields}.}
\label{fig.fieldperp}
\end{figure}

Let us close this section by pointing out the effect of the
second term in the energy expansion, Eq.~(\ref{DeltaE}), which is
proportional to $\uperp^2$. This term contains the influence of
the forces that the contacts carry before applying the external
point force, and is therefore referred to as the \emph{pre-stress}
term~\cite{alexander}. This term is strictly negative in a system
of purely repulsive particles, and therefore tends to destabilize
the system, to enhance the perpendicular (sliding) motions of
particles, and to lower the energy associated with deformations.
To study its effect, we also performed a calculation in which we left it out.
The destabilization can be seen from comparing
Fig.~\ref{fig.fieldperp}a with Fig.~\ref{fig.fieldperp}b: the
spatial fluctuations from the continuum elasticity stress fields
are much smaller if we leave out the pre-stress term. The fact
that the elastic energies are lower when the term $\propto
u_\perp^2$ is included in the energy expansion can also be shown
by calculating the elastic moduli with and without this pre-stress
term. The resulting shear modulus without this term is higher than
when this term is included. We will come back to the relative
importance of the two terms in the energy expansion in
section \ref{sectionenergyminimization}.

\section{Critical length scale}
\label{sec.lengthscale} In the
previous section we have seen that the ensemble averaged
response of a jammed granular medium can be described using
continuum elasticity. On the other hand, the disordered nature of
the packing has a strong effect on the displacements and forces in
individual realizations, especially in systems close to the
jamming transition. The question is
whether we can think about the role of disorder as relevant
on small length scales, but sufficiently smeared out on long
length scales. We will probe this question by locally forcing the
system, and studying the fluctuations away from the ensemble
averaged response as a function of length scale. We will find a
length scale $\ell^*$ which indicates above what length one can
consider the system a continuum medium. This length scale is
proportional to $1/\Delta z$, which implies that it diverges as we
approach the jamming transition.

\begin{figure}
\includegraphics[width=8.4cm]{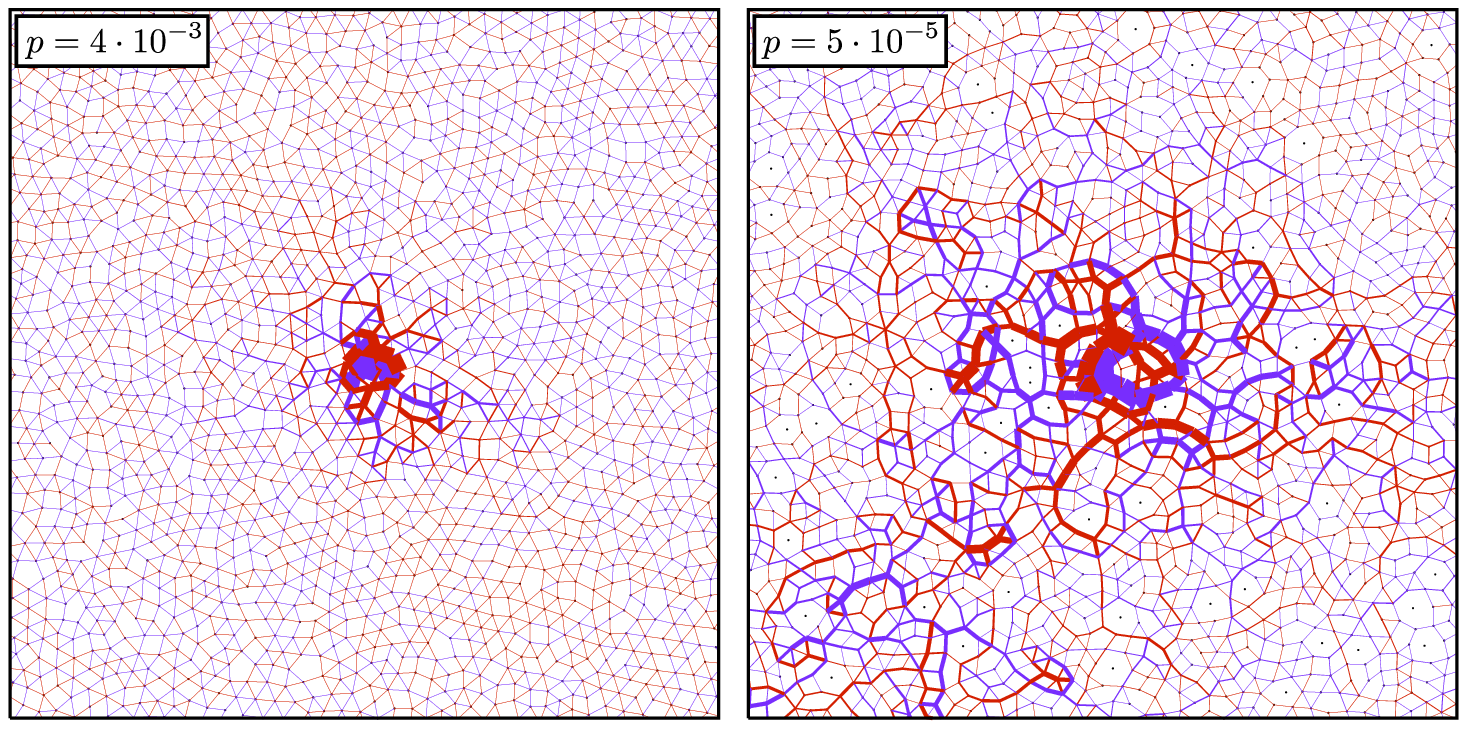}
\caption{(color) Force response networks for inflation of a single particle, with $p$
as indicated. Blue
(red) lines indicate positive (negative) changes in contact force, the
thickness indicating the amount, which is drawn for a 2\% increase of the particle diameter.
The thickness of the border around the panel corresponds to a change in force of 1/4500 (left), and 1/60000 (right),
respectively, which is needed because the shear modulus is much lower at the lower pressure.
The particles themselves are not drawn.
The figures show only the part of the packing close to the inflated grain,
and contain about 1800 of the 10\,000 particles.}
\label{fig:inflnetworks}
\end{figure}

The length  scale $\ell^*$ was introduced earlier by Wyart to
describe the excess of low frequency modes in the density of
vibrational states of these systems~\cite{wyartEPL}, where it can
be derived as follows.  If a system of dimension $d$ is to be
described as a continuum medium, it should keep its properties
when cut open and split in two parts. In particular, if we cut out
a circular blob of radius $\ell$ of a rigid material, it should
remain rigid. The rigidity (given by the shear modulus) of jammed
granular materials is proportional to $\Delta z$, the density of
excess contacts over the minimum number of contacts required to be
mechanically stable. The circular blob has of the order
$\ell^d\Delta z$ of such contacts. To cut it out, however, we had
to break the contacts at the perimeter, of which there are of
order $z\ell^{d-1}$. If the number of broken contacts at the edge
is larger than the number of excess contacts in the bulk, the
resulting blob is not rigid, but floppy (see
Appendix~\ref{app.floppy}). The smallest blob one can cut out
without it being floppy is obtained when these numbers are equal,
which implies that it has radius $\ell^*\sim z/\Delta z$.
Close to the jamming transition, $z$ is essentially constant, and
so one obtains as scaling relation that \cite{wyartPRE}
\begin{equation}
\label{deflstar} \ell^*\sim \frac{1}{\Delta z}~.
\end{equation}

So far this length scale $\ell^*$ has not been observed directly.
What has been observed is a crossover frequency $\omega^*$ in the
density of vibrational states, marking the lower end of a plateau
of excess states. The vibrational states at $\omega<\omega^*$ have
been interpreted as ordinary plane wave-like
modes~\cite{wyartEPL,silbertPRL05}. Using the speed of sound one
can therefore translate the crossover frequency into a wavelength,
which scales as $\lambda_\mathrm{T}\sim 1/\sqrt{\Delta z}$ for
transverse (shear) waves and as $\lambda_\mathrm{L}\sim 1/\Delta
z$ for longitudinal (compressional) waves. $\lambda_\mathrm{T}$
has been observed in the spatial structure of the vibrational
modes by Silbert \emph{et al.}~\cite{silbertPRL05}, but it is
$\lambda_\mathrm{L}$ that coincides with the length scale $\ell^*$
derived above. Note that the derivation of $\ell^*$
involved neither shear nor compression waves and therefore the
correspondence of $\ell^*$ and $\lambda_\mathrm{L}$ is not obvious
a priori. Below we present the first real-space observation of
$\ell^*$, in the fluctuations of the force response to a localized
perturbation. \newline

\subsection{Observation of $\ell^*$ in inflation response}
The signature of the length scale $\ell^*$ can readily be observed
in the point force response networks (see
Fig.~\ref{fig:respnetworks}): The lower the pressure, and hence
the smaller $\Delta z$, the larger the scale up to which the
response looks disordered. To study this effect quantitatively, we
calculate the response to an inflation of a single central
particle (Fig.~\ref{fig:inflnetworks}) instead of directional
point forcing (Fig.~\ref{fig:respnetworks}). This allows us to
probe a natural length scale of the system, as we expect a
crossover between the behavior for small and large $r$: far away
from the inflated particle we expect a smooth response with radial
symmetry, for which $\Delta\sigma_{rr}$ ~\footnote{See page 21 of
Ref.~\cite{lanlif7}.},
$$
\Delta\sigma_{rr}\approx\frac{G}{r^2}~,
$$
while nearby, the disorder dominates the response.  As we will
show now, the crossover length can be identified with $\ell^*$.

Examples of the changes in contact forces $\Delta f$ in
response to the inflation of a single particle are shown in
Fig.~\ref{fig:inflnetworks}, for pressures $p=4\cdot10^{-3}$ and
$p=5\cdot10^{-5}$.  For a fixed change in radius of the central
particle, the scale of $\Delta f$ is strongly influenced by the
stiffness of the few contacts of this particle, leading to large
fluctuations in the amplitude of $\Delta f$. Since we are mainly
interested in the spatial structure rather than the values of
$\Delta f$, we first normalize the force response $\Delta f$. To
do so, we fit the change of each local radial stress
$\Delta\sigma_{rr}$ to the continuum field (with a correction for
the periodic boundaries),
$$
\Delta\sigma_{rr}\approx\frac{G_\mathrm{fit}}{r^2}~,
$$
and define the normalized response $\Delta f^\mathrm{n}$ as $\Delta
f/G_\mathrm{fit}$.
In exceptional cases, the fitting parameter $G_\mathrm{fit}$ is
anomalously small. Inspection of the response networks in which
this happens reveals relatively large ``soft spots'', regions
where the rearrangements of the particles are large. We leave an
analysis of these soft spots and their relavance to the future;
for the present analysis we limit ourselves to noting that these
soft spots sometimes lead to very bad fits of the stress field,
and discard the 1\% worst-fitting response networks.

\begin{figure}
\includegraphics[width=8.4cm]{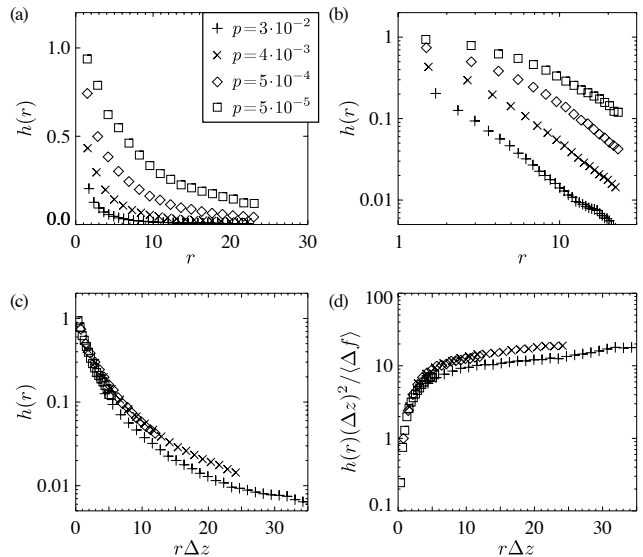}
\caption{Identification of the length scale $\ell^*$. (a)~The
fluctuations in the contact forces, measured by $h(r)$, which is
defined in equation~(\ref{defhfluc}), are larger in the response
of packings at low pressure. (b)~The same data on a double
logarithmic plot, showing that the tail decays as a power law,
approximately $1/r^{1.6}$. (c)~The data collapses when we rescale the
$r$-axis by $\Delta z$, signaling a length scale that scales as
$\ell^*\sim1/\Delta z$. (d)~The relative fluctuations
(fluctuations divided by the average value of the contact forces).}
\label{fig:lengthscale}
\end{figure}
In view of the radial symmetry, on average, we study the fluctuations in the response  at a
distance $r$ from the inflated grain. More precisely, we calculate the root mean
square fluctuations of the radial component of the normalized
force response, $\Delta f_r^\mathrm{n}$, through all contacts at that
particular distance $r$:
\begin{equation}
\label{defhfluc} h(r)\equiv\sqrt{\left\langle\left[\Delta
f_r^\mathrm{n}(r)-\langle \Delta f_r^\mathrm{n}(r)\rangle\right]^2\right\rangle}~.
\end{equation}
Here the average $\langle \cdot \rangle$ is taken over all bonds
that intersect the circle of radius $r$ centered around the
inflated particle. Note that $h(r)$ is not a simple correlation
function in the ordinary sense: it simultaneously involves
\emph{all} contacts at a distance $r$ from the center and cannot
be expressed in terms of a single $n$-point correlation function.
We calculate this function for packings of $N=10\,000$ particles
and average over 356 response networks at each value of $p$. The
resulting $h(r)$ are shown in Fig.~\ref{fig:lengthscale}a. As
expected, the fluctuations are larger and longer ranged for
packings at smaller pressure. Figure~\ref{fig:lengthscale}b shows
the decay on a double logarithmic plot: it appears that for large
$r$ the tail of $h(r)$ goes roughly as $1/r^{1.6}$ while the small
$r$ behavior has a smaller slope that varies with $p$. The
crossover scale between these regimes is proportional to the
length scale $\ell^*$, as is illustrated by the data collapse that
is obtained when $h(r)$ is plotted as function of $ r\Delta z$ in
Fig.~\ref{fig:lengthscale}c.
One might expect the fluctuations to decay as $r^{-2}$, following
the decay of the stress. We attribute the difference in exponent
to finite size effects.

In Fig.~\ref{fig:lengthscale}d we plot the \emph{relative}
fluctuations with respect to the average radially transmitted
force $\langle\Delta f_r^\mathrm{n}(r)\rangle$
(Fig.~\ref{fig:lengthscale}d). These are nearly constant ($\sim
r^{0.4}$) after $r\Delta z\approx 6$, which can serve as a choice
of prefactor for the crossover length:
\begin{equation}\label{ellexpression}
\ell^*\approx\frac{6}{\Delta z}~.
\end{equation}
For large $r$ the relative fluctuations do not go to zero, which
indicates the system is not self-averaging. The length scale
$\ell^*$ thus marks the distance at which the relative error stops
growing. Note again that the $r$ in this analysis refers to the
distance from the perturbation where the flucuations are measured,
and that the fluctuation analysis does not involve a coarse
graining length. The fluctutations, at any $r$, are measured on
the scale of single contacts. This contrasts with the next
subsection, where we will relate $\ell^*$ to the coarse graining
length needed to obtain a smooth response.

\subsection{Inhomogeneity of global response}
\label{subsec.lstarglob}
\begin{figure}[!bt]
\includegraphics[width=8.4cm]{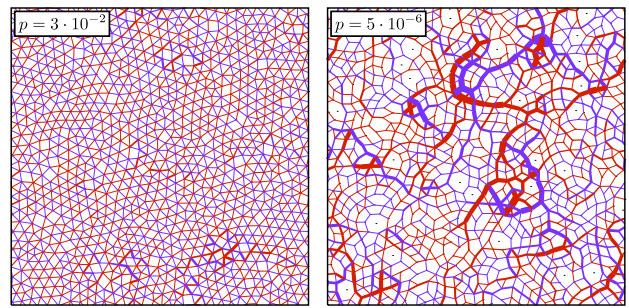}
\caption{(color) Force response networks for a global shear deformation of
packings of 1000 particles at high (left) and low (right) pressure.
The thickness of the border around the panel corresponds to a change in force
of 0.5 (left), and 80 (right) per unit strain, respectively, which is needed because
the shear modulus is much lower at the lower pressure.
Note that in the high pressure case nearly all bonds that have an angle between 0
and $\pi/2$ with the $x$-axis are red (decreased force) and nearly all bonds
in the other diagonal direction are blue (increased force), consistent with
the observation that the displacement fields for this shear response are
very close to affine (Fig.~\ref{fig:dispfield}). The response network of
the low pressure packing is much more disordered.}
\label{fig:shearrespnetworks}
\end{figure}
\begin{figure}[!bt]
\includegraphics[width=8.4cm]{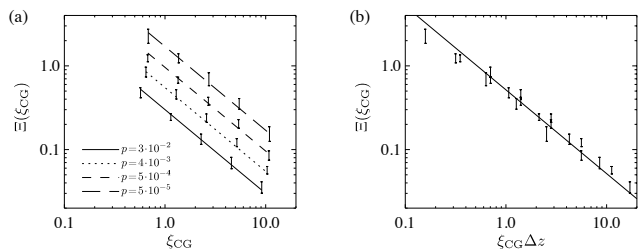}
\caption{(a) Inhomogeneity $\Xi$ of the response as a function of
coarse graining length $\xi_\mathrm{CG}$, for various pressures.
The lines indicate the $1/\xi_\mathrm{CG}$-behavior and the error
bars denote the $20\%$ median values of our dataset of 90 packings
at each pressure. (b) The same data, with the horizontal axis
rescaled by $\Delta z$ to collapse the data. The line again
denotes $\Xi\sim(\xi_\mathrm{CG}\Delta z)^{-1}$.}
\label{fig:cgshear}
\end{figure}

The results for the crossover length suggest that to obtain a
smooth stress response field by coarse graining $\Delta f$, one
would need to use a coarse graining length proportional to
$\ell^*$. To test this explicitly, we now study the stress
fluctuations in packings under an overall shear as a function of
the coarse graining length $\xi_\mathrm{CG}$.

The starting point are force response networks for packings
of $10\,000$ particles that we obtain in linear response to shear
(Fig.~\ref{fig:shearrespnetworks}).
The behavior is consistent
with what we
observed for point forcing: the characteristic length scale of
these force fluctuations appears to grow when approaching the
jamming point. To quantify this, we calculate the change in
shear stress $\Delta\sigma_{xy}$ in 64 points using
equation~(\ref{stresscoord}) with a Gaussian coarse graining
function. The (relative) standard deviation of the resulting 64
values is a measure of the inhomogeneity of the response:
$$
\Xi(\xi_\mathrm{CG})=\frac{1}{\langle\Delta\sigma_{xy}\rangle}\sqrt{\left\langle\left(\Delta\sigma_{xy}(x_i)-\langle\Delta\sigma_{xy}\rangle\right)^2\right\rangle}~,
$$
where the averages are taken over the 64 points within each
separate packing. In Fig.~\ref{fig:cgshear} we plot
$\Xi(\xi_\mathrm{CG})$ and obtain that the inhomogeneities decay
as $1/\xi_\mathrm{CG}$. The force fluctuations thus lack an
intrinsic length scale. We can however collapse the data for $\Xi$
for different pressures when they are plotted as function of
$\xi_\mathrm{CG} \Delta z$. Therefore, the length-scale $\ell^*\sim 1/\Delta z$,
which was observed in the inflation response in the previous
subsection, can also be used to set the coarse graining length one
needs to obtain a stress field with a particular inhomogeneity
$\Xi$.
It should be noted that Fig.~\ref{fig:cgshear} also implies that, for a fixed coarse graining
length, the ensemble size one would need to obtain a desired smoothness of
response grows as $(\Delta z)^{-2}$.

\section{Energy minimization and local sliding}
\label{sectionenergyminimization}

In this section we will explore various connections between the local
displacement field, global energy minimization, and the scaling of
the elastic moduli. In section~\ref{subsec.simple} we present
simple scaling arguments for the typical magnitude of $\uperp$,
$\uparl$ and their ratio as function of the distance to the
jamming point. In section \ref{subsec.alpha} we employ the
displacement angle distribution $P(\alpha)$ \cite{respprl} to
verify the latter scaling prediction, and find that both under
shear and under compression, the response of jammed packings
becomes increasingly non-affine near the jamming point. In section
\ref{subsec.udiverge} we test the scaling
predictions for the probability distributions $P(\uparl)$ and
$P(\uperp)$, and find that $\uperp$ diverges near the jamming
point, as predicted. However, while our simple scaling arguments
capture the behavior of systems under shear in detail, they do not
quantitatively capture the scaling of $P(\uparl)$ and $P(\uperp)$
for jammed packings under compression.

\subsection{Simple scaling arguments}\label{subsec.simple}
We now present a set of simple scaling relations connecting the local
displacement field to the scaling of the elastic moduli and global energy
minimization.
The starting point for our discussion is the energy expansion,
equation~(\ref{DeltaE}), which is a function of the relative
displacements of particles in contact, and which for our power law
interparticle potential reads \footnote{For a general potential
$V_{ij} \sim \delta_{ij}^\alpha $ the factor 2/3 in
(\ref{DeltaE2}) becomes $1/(\alpha -1)$.}:
\begin{equation}
\label{DeltaE2} \Delta E=\frac12\sum_{\langle i,j\rangle}
k_{ij}\left(u_{\parallel,ij}^2-\frac23\delta_{ij}u_{\perp,ij}^2\right)~.
\end{equation}
As explained in section~\ref{subsec.linresp}, $k_{ij}$ is the
stiffness $\tdiff{^2 V_{ij}}{r_{ij}^2}$, where $V_{ij}$ denotes
the interparticle potential, $\delta_{ij}$ is the dimensionless
overlap $1-r_{ij}/(R_i+R_j)$, and $\uparl$ ($\uperp$) denotes the
parallel (perpendicular) component of the relative displacement
$\vc{u}_{ij}$, with respect to the vector $\vc{r}_{ij}$ connecting
the centers of the contacting particles.

For the response to a global shear or compression, $\Delta E$ is also related to the
elastic moduli, through
\begin{eqnarray}
\Delta E\sim C\epsilon^2&\sim&\frac12\sum_{\langle i,j\rangle}
k_{ij}\left(u_{\parallel,ij}^2-\frac23\delta_{ij}u_{\perp,ij}^2\right)~,
\end{eqnarray}
where $C$ refers to $G$ and $K$ for shear and compressive deformations,
respectively, and $\epsilon$ is the applied strain.
Note that the first term between brackets is strictly positive and the second term
is strictly negative. Because the packings were constructed to be at an energy
minimum, this implies that the second term cannot dominate over the first one, and hence the scaling
of typical values of $\uparl$ is connected to that of the corresponding modulus.
Inserting $K/k\sim\delta^0$ and $G/k\sim\delta^{1/2}$~\cite{epitome}, we obtain
\begin{eqnarray}
\uparl&\sim&\epsilon \, \delta^{1/4}\qquad\mathrm{for~shear}\label{uparlshearpred}\\
\uparl&\sim&\epsilon \, \delta^{0}\qquad\mathrm{for~compression,}\label{uparlcomppred}
\end{eqnarray}
where symbols without ${ij}$-indices refer to typical or average
values of the respective quantities. Note that by dividing out the stiffness
$k$ we obtained exponents that are valid for both harmonic and hertzian
interactions, and in fact should be valid for all potentials of the form
$V\sim\delta_{ij}^\alpha$.

The expected scaling for $\uperp$, the amount by which particles in contact
slide past each other, is more subtle. Because of the negative prefactor
in Eq.~(\ref{DeltaE2}), energy minimization should maximize these $\uperp$.
Again, they are bounded by the fact that these packings are stable:
$u_{\parallel,ij}$ and $u_{\perp,ij}$ are not the independent variables of the
problem, as they are coupled through the packing geometry, and for stable packings
the question really is how close to the typical $\uparl^2$ the typical $\delta\uperp^2$
can get. The best the system could do in order to minimize the change in energy is to make them
of the same order, which suggests that
\begin{equation}\label{scale_prediction1}
\frac{\uparl}{\uperp}\sim\sqrt\delta~.
\end{equation}
In particular, this means that the system is always close to a buckling instability, so that
compressing it would inevitably lead to the formation of more contacts to restabilize the
system~\cite{wyartPRE}. For the displacements in low energy eigenmodes,
Eq.~(\ref{scale_prediction1}) has recently been derived~\cite{wyart08} --- more
intuitively (\ref{scale_prediction1}) can be understood as arising from the
elastic distortion of floppy modes on a scale $\ell^*\sim1/ \Delta z \sim
1/\sqrt{\delta}$ \cite{wouterEPL09}. Generically, one may expect this property
to carry over to the displacements in response to external perturbations, but
this is by no means guaranteed.  Indeed, we shall find that
Eq.~(\ref{scale_prediction1}) and the corresponding predictions for the scaling
of $\uperp$ break down in case of a global compression, while they work very
well for the shear response. In a related paper, we studied this issue in
depth in the context of networks of harmonic springs~\cite{wouterEPL09}.

It is important to understand the relation between Eq.~(\ref{DeltaE2}) and the
excess coordination number $\Delta z=z-z_\mathrm{iso}$. First, note that for
$z<z_\mathrm{iso}$, deformation fields exist for which $\Delta E =0$ --- the
so called floppy modes (see Appendix \ref{app.floppy}). Generically, right at
the jamming transition ($z=z_\mathrm{iso}$), there are just enough degrees of
freedom to set all individual $u_{\parallel,ij}$ equal to zero for the type of
response considered, in other words, to have sideways sliding motion for all of
the bonds at jamming.  For hyperstatic packings with $z > z_\mathrm{iso}$, the $\uparl$ cannot
all be set to zero anymore, because there are more of them than the number of degrees of freedom
$dN$. In addition, the $N$ $d$-dimensional coordinates are sufficiently constrained so that
displacements which make the negative terms in $\Delta E$ outbalance the
positive ones are forbidden, and $\Delta E>0$.

A thorough analysis of (\ref{scale_prediction1}) involves
the scaling of the distribution of the ratios
$u_{\parallel,ij}/u_{\perp,ij}$, since the average of $\uperp$ and
$\uparl$ over all contacts may be dominated by large and rare
fluctuations. To quantify the relative amount of sliding and
deformation, we therefore introduce the displacement angle
$\alpha_{ij}$, defined as the angle between $\vc{u}_{ij}$ and
$\vc{r}_{ij}$, or,
\begin{equation}
\label{defalpha}
\tan\alpha_{ij}=\frac{u_{\perp,ij}}{u_{\parallel,ij}}~.
\end{equation}
In section~\ref{subsec.alpha} below we will present the
probability densities $P(\alpha)$ for shear and compressive
deformations as a function of
the distance to point J, and test the prediction given by
Eq.~(\ref{scale_prediction1}) in detail.

\begin{figure*}
\includegraphics[width=\textwidth]{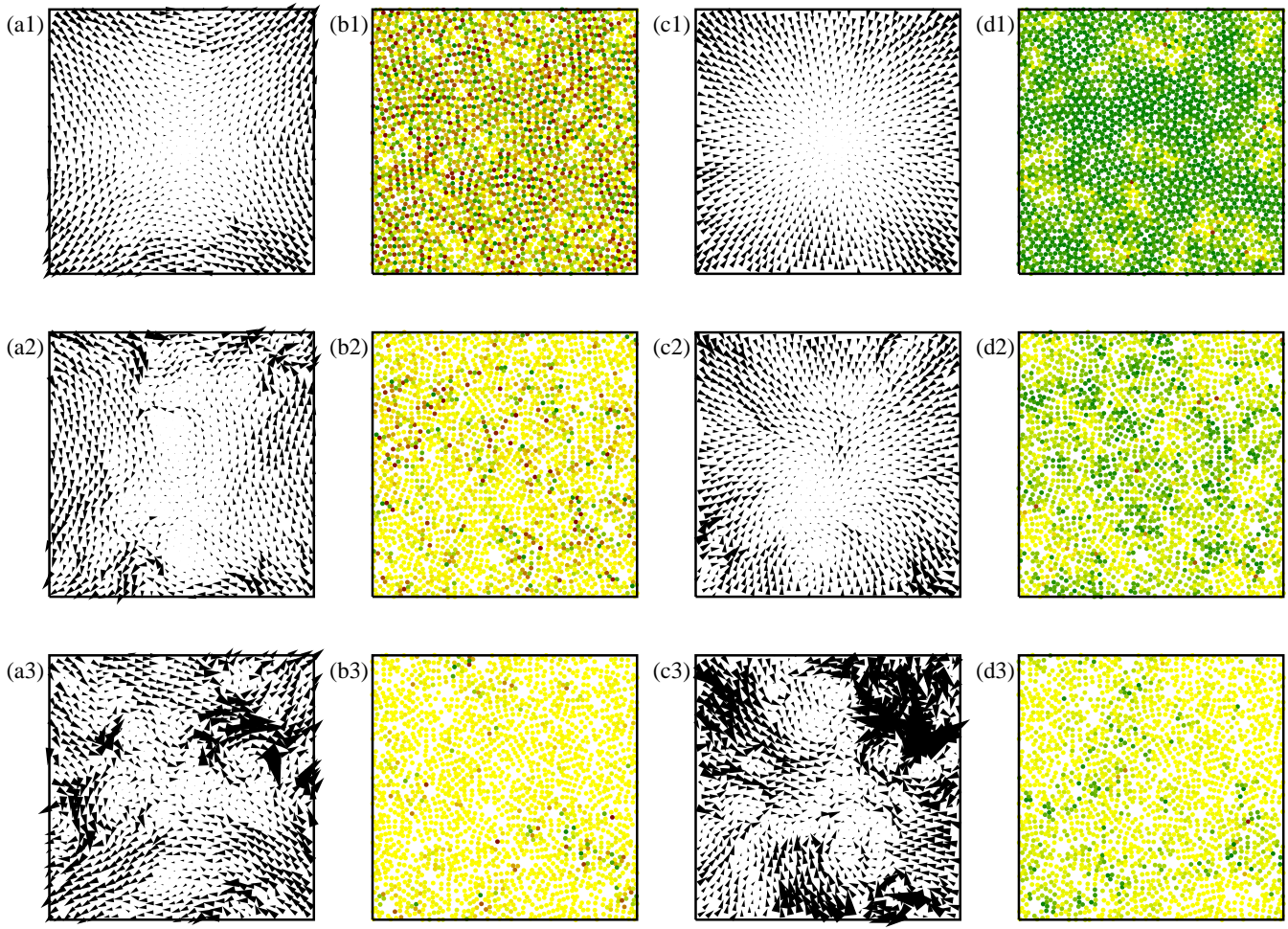}
\caption{(color online) Examples of displacement fields in linear
response to shear (a) and compression (c) for three pressures,
$p=3\cdot10^{-2},5\cdot10^{-4},5\cdot10^{-6}$, from top to bottom.
In (b) and (d) the corresponding displacement angles are depicted,
where bonds are marked with a dot, on a color scale which runs
from red ($\alpha=0$), via yellow ($\alpha=\pi/2$) to green
($\alpha=\pi$) --- see Eq.~(\ref{defalpha}) for the definition of
$\alpha$. Clearly, the highest pressure packing displays almost
affine displacements, while at lower pressures the response
becomes increasingly non-affine.} \label{fig:dispfield}
\end{figure*}

In section~\ref{subsec.udiverge} we will study the scaling of the
distributions of $\uparl$ to test the predictions of Eqs.~(\ref{uparlshearpred},\ref{uparlcomppred}).
We will also present the the distributions of $\uperp$ under shear and compression, and explore to what
extent these saturate the stability bound given by $\uperp\sim\uparl/\sqrt\delta$. As alluded to before,
we will show that for the response to shear, $\uperp$ follows the prediction from combining
Eqs.~(\ref{uparlshearpred}) and (\ref{scale_prediction1}), namely
\begin{eqnarray}
\uperp&\sim&\epsilon \, \delta^{-1/4}\qquad\mathrm{for~shear.}\label{uperpshearpred}
\end{eqnarray}
The corresponding scaling for the response to compression would be, from Eqs.~(\ref{uparlcomppred}) and
(\ref{scale_prediction1}),
\begin{eqnarray}
\uperp&\sim&\epsilon \, \delta^{-1/2}\qquad\mathrm{for~compression,}\label{uperpcomppred}
\end{eqnarray}
but we will show that the numerical results do not support this --- in the case of compression the
two terms in the expansion of $\Delta E$ do not become of the same order.

Note that Eq.~(\ref{scale_prediction1}) predicts that the response
of the system becomes strongly non-affine near point J, and that
Eqs.~(\ref{uperpshearpred},\ref{uperpcomppred}) predict that typical
distances by which particles slide past each other \emph{diverge}
as we approach the jamming transition. In finite systems, this divergence of the
local displacements is limited by finite size effects. We discuss the corresponding
finite size scaling and crossover in Appendix~\ref{subsec.finite}.

\subsection{The displacement field and displacement angle distribution}
\label{subsec.alpha} We extract the displacement fields from our
linear response calculations of the periodic packings of 1000
particles. Examples of these, for both compression and shear and
for three different pressures, are shown in
Fig.~\ref{fig:dispfield}a,c, and the corresponding displacement
angles $\alpha$ are shown in Fig.~\ref{fig:dispfield}b,d.

For high pressure, far from point J, the displacement fields are
quite smooth and close to affine deformations, i.e.,
$\vc{u}=\epsilon(y\uvc{x}+x\uvc{y})$ for shear and
$\vc{u}=-\epsilon(x\uvc{x}+y\uvc{y})$ for compression. When point
J is approached, the displacement fields clearly become more
disordered. The displacements become increasingly non-affine, and
organize in  vortex-like structures. Similar structures have been
observed in various experimental and numerical studies of
disordered systems close to the jamming
transition~\cite{granulence,tanguy,ellakwave,lemaitre06,tanguy04,maloneyPRL06,
blairprivate}.

In principle, this non-affine behavior can be analyzed by
subtracting the appropriate affine displacement field  from the
observed displacement field
\cite{leonforte,tanguy04,maloneyPRL06}.  By doing so, however,
valuable information on the correlation between $\rij$ and $\uij$
is lost. Therefore, we analyze the nature of the non-affinity in
terms of the displacement angles $(\alpha)$, where $\alpha$ is
defined via
$\tan\alpha_{ij}=\frac{u_{\perp,ij}}{u_{\parallel,ij}}$ (see
Eq.~(\ref{defalpha})). Under an affine compression, all particles
get closer together along their contact line and $P(\alpha)$ would
be a $\delta$-peak at $\alpha=\pi$. For a purely affine shear
displacement field in an isotropic system, $P(\alpha)$ is
flat~\footnote{This is easily seen in the case of a simple shear
in the $x$-direction: All $\vc{u}_i$ are in the $x$-direction,
hence so are the $\uij$, while the $\rij$ are isotropic.}.
Finally, close to point J where $\delta \rightarrow 0$,
Eq.~(\ref{scale_prediction1}) predicts that
${u_{\perp,ij}}/{u_{\parallel,ij}}$ diverges, i.e., that
$P(\alpha)$ approaches a $\delta$-peak at $\alpha=\pi/2$
(see also Appendix~\ref{subsec.finite}).

The values of $\alpha$ in Figs.~\ref{fig:dispfield}b1 and
\ref{fig:dispfield}d1 are what is expected for an affine
displacement. One sees that close to jamming, more and more
bonds have $\alpha$ close to $\pi/2$ (corresponding to the bright
yellow dots in Figs.~\ref{fig:dispfield}b2 and
\ref{fig:dispfield}d2 and in particular Figs.~\ref{fig:dispfield}b3
and \ref{fig:dispfield}d3), indicating that for those displacement
fields $\uparl\ll\uperp$, as predicted in
Eq.~(\ref{scale_prediction1}).

\begin{figure}
\includegraphics[width=8.4cm]{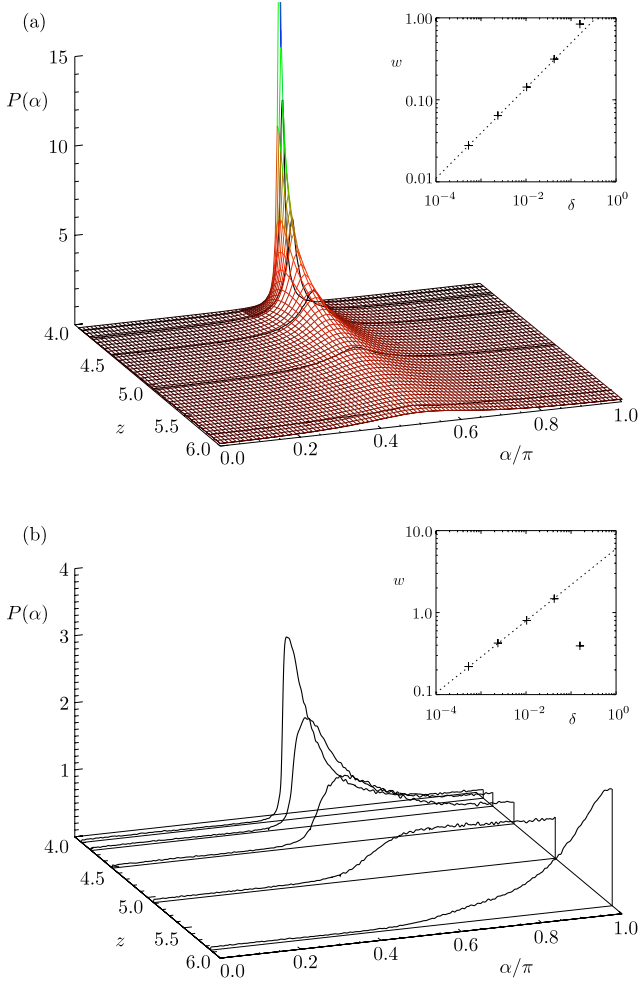}\\
\caption{(color online) The displacement angle distribution for various
pressures, obtained by averaging over 100 packings for each
pressure. (a) Under shear $P(\alpha)$ evolves from nearly flat
(highest pressure) to very sharply peaked at $\alpha=\pi/2$
(lowest pressure). The solid lines represent numerical data, the
gridded surface Eq.~(\ref{pashearfit}). Inset: The scaling of the
width of the peak as a function of typical overlap $\delta$. The
dashed line indicates $w\sim\delta^{0.55}$.
(b) Under compression $P(\alpha)$ has a peak at $\pi$ for high
pressures and develops a peak at $\pi/2$ at low pressures. Inset:
The scaling of the width of the peak as a function of typical
overlap $\delta$. The dashed line indicates $w\sim\delta^{0.44}$.
} \label{fig:palpha}
\end{figure}

For the rest of the paper we will ignore the spatial organization
of the displacement angles $\alpha$, and focus on their
probability densities $P(\alpha)$. In Fig.~\ref{fig:palpha} we
show these probability distributions for compressional and shear
deformations for a range of pressures. For large pressures,
$P(\alpha)$ reflects the affine limits discussed above, while upon
lowering the pressure, $P(\alpha)$ evolves to exhibit a strong
peak at $\alpha=\pi/2$, as predicted by
Eq.~(\ref{scale_prediction1}). In the case of the shear
deformation, we find that the peak in $P(\alpha)$ can be well
approximated by a Lorentzian, and we fit $P(\alpha)$ to~\footnote{Because of
the $\pi$-periodic nature of the angle $\alpha$, we add copies of the peak at
$-\pi/2$ and at $3\pi/2$. This improves the fit for larger pressures.}
\begin{equation}
\label{pashearfit} P(\alpha)\simeq
\frac{1}{\pi\,w} \,\frac{1}{1+\left(\frac{\alpha-\pi/2}{w}\right)^2}~.
\end{equation}
Note that the width $w$ should vary as $\uparl/\uperp$ close to the jamming
transition, because $|\alpha_{ij}-\pi/2|\approx
u_{\parallel,ij}/u_{\perp,ij}$ if $u_{\parallel,ij}\ll
u_{\perp,ij}$

In good approximation, the resulting widths satisfy the scaling
relation
\begin{equation}
\label{widthscale}
w \sim \delta^{0.55\pm0.05}~,
\end{equation}
as is shown in the inset of Fig.~\ref{fig:palpha}a. This scaling
has been used to construct the colored surface in
Fig.~\ref{fig:palpha}a from Eq.~(\ref{pashearfit}), and is
consistent with the prediction of Eq.~(\ref{scale_prediction1}),
that $w \sim \sqrt{\delta}$.

For the case of compression the situation is more complex as
can be seen from Fig.~\ref{fig:palpha}b. Since the distribution
here does not appear to be governed by a single scale and we do
not have a fitting function, we have used the full width at half
maximum of $P(\alpha)$ as $w$. The value of this width is shown in
the inset of Fig.~\ref{fig:palpha}b. It scales as $w\sim
\delta^{0.44[5]}$, which is reasonably consistent with
Eq.~(\ref{scale_prediction1}). However, the presence of a clear
shoulder for $\alpha\simeq \pi$ is a strong indication a simple
one-parameter scaling actually does not hold \cite{wouterEPL09}.

The shear data shown in Fig.~\ref{fig:palpha} are thus in agreement with
the prediction from the simple scaling argument that near point J,
$\uparl/\uperp\sim\sqrt\delta$
(Eq.~(\ref{scale_prediction1})). Hence, we conclude that the displacements in
response to shear do lead to an approximate balance between the two terms in
$\Delta E$ (Eq.~(\ref{DeltaE2})). The closer one gets to point J, the more
strongly non-affine the deformation field becomes, in other words, the more the
mechanical response of the system is different from that of a homogeneous,
isotropic elastic material. For compressive deformations, the situation is
somewhat more complicated: the development of a peak at $\alpha\approx \pi/2$
still signifies the same tendency to non-affine sideways sliding motion with an
amplitude that might be described by a balance of the two terms in the energy expression,
but the remainder of a significant shoulder at large angles $\alpha$ hints at
the fact that the response can not be understood completely in terms of this balance or a simple one-parameter scaling.
For more details, see Ref.~\cite{wouterEPL09}.

\subsection{Scaling of local displacements}
\label{subsec.udiverge} As we discussed in the introduction to this section, by
linking the scalings of the local relative displacements to the elastic moduli,
we can predict scaling relations for $\uparl$, and, when the two terms in the
energy expansion are of the same order, for $\uperp$, both for compressive and
shear deformations. We now test these predictions by plotting the distributions
of the rescaled displacements $P(\uparl \delta^{n_{\parallel}})$ and $P(\uperp
\delta^{n_{\perp}})$, where the power law indices $n$ follow from
Eqs.~(\ref{uparlshearpred},\ref{uparlcomppred},\ref{uperpshearpred},\ref{uperpcomppred}),
and check for data collapse.

\begin{figure}
\includegraphics[width=8.4cm]{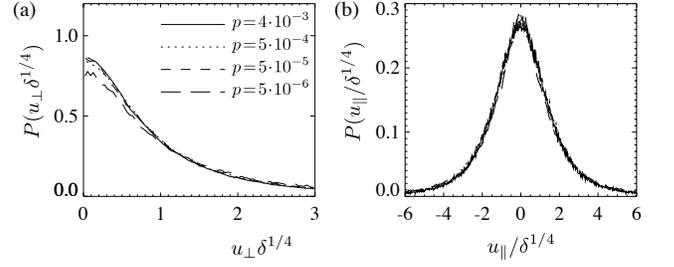}
\caption{Collapse of the probability densities of $\uperp$ (a) and
$\uparl$ (b) for a shear deformation. The axes have been rescaled
according to the prediction of Eq.~(\ref{uparlshearpred}) and (\ref{uperpshearpred}).
Higher pressures have been omitted from the analysis.}
\label{fig:pushear}
\end{figure}

\begin{figure}
\includegraphics[width=8.4cm]{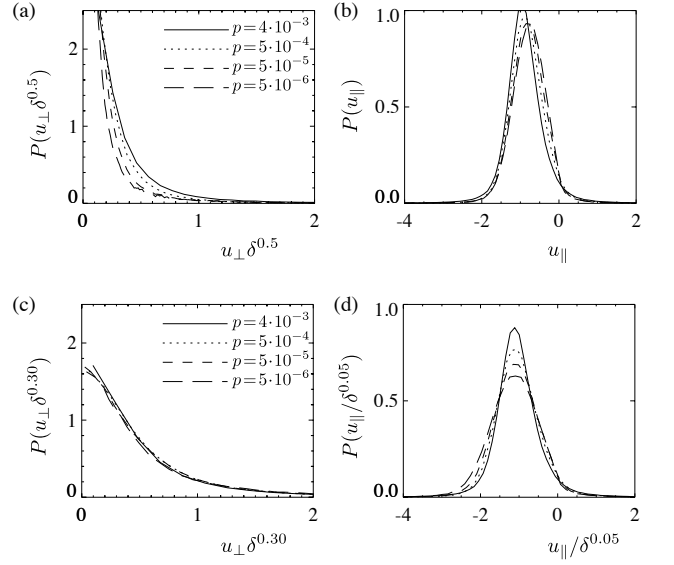}
\caption{Rescaled probability
distributions of $\uperp$ and $\uparl$ for compressional
deformations. (a-b) Rescaling of $P(\uperp)$ and  $P(\uparl)$ with
the exponents predicted by Eqs.~(\ref{uparlcomppred}) and (\ref{uperpcomppred}) produces a
poor collapse, indicative of the special nature of compression of
jammed packings. (c-d) When the scaling exponents are adjusted by
hand, from 0.5 to 0.3 (c), and from 0 to 0.05 (d), a
reasonable collapse can be obtained. As before, in all
panels the distribitions for the highest pressure have been
omitted.}
\label{fig:pucomp}
\end{figure}

The appropriately rescaled distributions of $\uparl$ and $\uperp$ are shown in
Fig.~\ref{fig:pushear} for the case of shear deformations. Note that we
normalize the displacements by the strain $\epsilon$~\footnote{In addition, it
is natural to eliminate systematic dependence of $\uparl$ and $\uperp$ on the
particle radii, and we have divided each $\uperpij$ and $\uparlij$ by the
actual length $r_{ij}$ of the contact --- this would also allow to compare the
results to the case of affine displacements, which have local $\uperpij$ and
$\uparlij$ which are proportional to $r_{ij}$, but we have already seen that
the displacements are strongly non-affine, so we do not show the corresponding
plot here.}. As is clear from Fig.~\ref{fig:pushear}, the scaling collapse is
very good for both components of the displacement field $\uij$, fully
confirming the simple prediction for $\uparl$ presented in
Eq.~(\ref{uparlshearpred}), and the prediction for $\uperp$ in
Eq.~(\ref{uperpshearpred}) that follows from the assumption that the two terms in the
energy expansion are of the same order.

For the displacements under compression we show scaling
collapses in Fig.~\ref{fig:pucomp} \footnote{Note that because the
distribution of $\uparl$ under compression is not centered around
zero, we are collapsing the average of the distributions, not the
width.}. In Fig.~\ref{fig:pucomp}a-b we show the distributions
$P(\uperp \delta^{0.5})$ and $P(\uparl \delta^0)$, which
Eqs.~(\ref{uparlcomppred}) and (\ref{uperpcomppred}) predict to collapse. Clearly, the collapse
is far worse than in the case of jammed packings under shear.
Similar to $P(\alpha)$, the distributions $P(\uperp)$ and
$P(\uparl)$ do not appear to exhibit simple scaling for jammed
packings responding to compression. The best scaling collapse is
obtained by adjusting the scaling exponents $n$ by hand, but even
then, Fig.~\ref{fig:pucomp}c-d show, that this adjusted scaling
collapse is still less convincing than for systems under shear.
The exponent
$n_\perp=0.30$ is significantly different form the predicted value of 0.5, reflecting that
the displacements in response to compression satisfy neither
a balance of terms in the energy expansion nor pure one-parameter scaling \cite{wouterEPL09}.

In conclusion, the displacement angle distribution shows the development of a
peak in $P(\alpha)$ at $\alpha =\pi/2$, both for shear and compressive
deformations. The width of the peak shrinks as $\sqrt{\delta}$, consistent with
the energy balance argument that predicts that $\uparl/\uperp ~ \sqrt{\delta}$.
This peak signals that near point J, most particles in contact mainly slide
past each other, in a manner which helps to minimize the changes in elastic
energy. However, in the case of compression $P(\alpha)$ retains a significant
shoulder close to $\alpha=\pi$, even close to the jamming transition.

Both for shear and compressive deformations, the sliding motion
$\uperp$ diverges near jamming. This suggests that corrections due
to finite size play a role --- for details see
Appendix~\ref{subsec.finite}. For shear deformations, simple
arguments predict the scaling of $P(\uparl)$ and $P(\uperp)$
quantitavily correct, but for compressive deformations our simple
arguments break down. We believe that this fact can ultimately be traced back to the special
geometry that the contact network of a packing has because it is made of
particles that interact purely repulsively. We have recently studied this question
in the context of harmonic networks --- for more information see Ref.~\cite{wouterEPL09}.

\section{Summary and Outlook}
We have shown, by means of linear response calculations, how
the applicability of elasticity theory to disordered packings of
frictionless spheres is related to the distance to the jamming
transition.
Averaging the response to a local perturbation over an ensemble of packings, we fitted
the stress response to continuum elasticity.
The resulting elastic moduli are the same as those obtained by
calculating the response to a global shear or compression, proving the consistent
applicability of continuum elasticity in an ensemble averaged sense.
In \emph{single packings}, we identified a length scale
$\ell^*\sim 1/(\Delta z)$ up to which the response is dominated by local disorder.
Since $\ell^*$
diverges as the jamming transition is approached, continuum elasticity breaks down completely as a description
of the linear response of individual packings near jamming.
The length
scale $\ell^*$ corresponds to the length used by Wyart \emph{et al.}\ to estimate the number
of soft modes in systems near jamming~\cite{wyartEPL}. Here we have shown that it also
represents the coarse graining length needed to obtain a smooth stress response
tensor in a single globally deformed packing.

The displacement fields near jamming are highly non-affine. We analyzed
these displacements at the grain scale through the statistics of local relative
displacements of neighboring particles.
These relative displacements, decomposed into $\uparl$ along the line
of contact, and $\uperp$ perpendicular to it, govern the elastic behavior
of the system because according to Eq.~(\ref{DeltaE}) they enter into the
change in energy as
$$
\Delta E=\frac12\sum_{\langle i,j\rangle}
k_{ij}\left(u_{\parallel,ij}^2-\frac{f_{ij}}{k_{ij}r_{ij}}u_{\perp,ij}^2\right)~.
$$
We have introduced the displacement angle distribution $P(\alpha)$ to easily
characterize the non-affinities, via the angle between the relative
displacement vector and the vector connecting two neighboring particles. Close
to point J, these distributions develop a peak near $\alpha=\pi/2$, indicating
that the non-affinity is such that particles predominantly slide past each
other.

We also
determined the scaling of typical values of $\uperp$ and $\uparl$ separately.
The parallel displacements $\uparl$ follow a scaling that is consistent with
the scaling of the elastic moduli: they are nearly independent of distance to
jamming for compression and vanish as $\delta^{1/4}$ for shear. The
perpendicular displacements $\uperp$ diverge in response to both shear and
compression, consistent with the predominance of sliding near jamming.

There is a surprising and fundamental difference between compression and shear.
For shear, $\uperp$ simply scales as $\delta^{-1/4}$, so that both terms in the
elastic energy are of the same order, and the scaling of $P(\alpha)$ is
captured by a single parameter. For bulk compression, all of this breaks down
--- $\uperp$ still diverges, but not with an exponent that follows from a
simple scaling argument, and there is no one-parameter scaling for $P(\alpha)$.
The issue of what is special about the compression of packings is discussed in
a separate paper~\cite{wouterEPL09}.

There are many extensions of this framework beyond the linear
response of frictionless discs in a computer simulation, of which
the extension to three-dimensional spheres is probably the
simplest. For finite displacements, the direct connection between
displacements and energy is lost. However, in any system of
approximately spherical particles close to the jamming transition,
where steric hindrance becomes an important factor in the
dynamics, one would expect the displacement angle distribution to
develop a maximum near $\alpha=\pi/2$, corresponding to
predominantly sliding displacements. These local displacement
statistics are accessible experimentally, even outside the linear
response regime, for example in steadily sheared (wet) foams or
emulsions, and even in systems below the jamming density, for
example in driven granular gases~\cite{abatePRE06,
reisPRL07,reisPRE07,keysNP07}. In the latter case, one has to
define neighbors (e.g. through a Voronoi tesselation), and one has
to choose the time scale $\tau$ over which to measure the relative
displacements so that neighboring particles have had the time to
interact --- $P(\alpha )$ will depend on $\tau$, and preliminary
studies suggest that the peak in $P(\alpha )$ is maximal for a
finite value of $\tau$. Note that in dense foams or emulsions, in
which the particles are continuously interacting, one rather
expects the peak of $P(\alpha )$ to be maximal for $\tau
\rightarrow 0$.

In systems of
ellipsoids~\cite{donev04,wouterse07,zorana09,bulbul09} or
frictional particles~\cite{maksePRL,ellakjam}, rotations and
torques come into play, and the expression for the elastic energy
will be more complicated than Eq.~(\ref{DeltaE}).  However, in
linear response the elastic energy will still be a function of the
local relative changes of the particle coordinates, and one can
still check for the occurrence of combinations of local
displacements that lead to a small or vanishing change in energy.
For example, in packings of frictional spheres, the locally floppy
displacement is for contacting particles to roll without
slipping. For ellipses ``zero energy'' local
displacements can be identified similarly. We expect that such low energy
displacements should dominate the statistics when the system is
close to the relevant isostatic limit, and we expect that
appropriate generalizations of $P(\alpha)$ for such systems may
serve as a useful characterization of the non-affine deformation
field.

\begin{acknowledgments}
We thank E. Somfai for providing the numerical code to construct
the granular packings, and for many crucial discussions. In
addition, we are grateful to M. Wyart for several discussions that
have been very enlightening and helpful. We wish to thank N. Xu,
K. Shundyak, J.H. Snoeijer, M. Depken, C. Goldenberg, S. Ostojic,
S. R. Nagel, A. J. Liu, Z. Zeravcic, and H. van der Vorst for useful
discussions and suggestions. WGE acknowledges support from the
Stichting voor Fundamenteel Onderzoek der Materie (FOM), and MvH
from the Dutch science foundation NWO through a VIDI grant. WGE
thanks the Aspen Center for Physics, where part of this work was
done, for its hospitality.
\end{acknowledgments}

\appendix

\section{Packing generation}
\label{app.packgen}

\subsection{The Hertzian interaction}
\label{app.Hertz}
The Hertzian interaction is taken from Refs.~\cite{contact,ellakwave} in
terms of a \emph{force law} as
\begin{equation}
\label{HertzJohnson}
f_{ij}=\frac43\sqrt{R_{ij}}E^*_{ij}\left(R_i+R_j-r_{ij}\right)^{3/2}~,
\end{equation}
where $R_{ij}$ is the reduced radius of the pair of particles, defined by
$$
\frac1{R_{ij}}=\frac1{R_i}+\frac1{R_j}~,
$$
and $E^*_{ij}$ is a similar combination of the modified Young moduli,
$$
\frac1{E^*_{ij}}=\frac1{E^*_i}+\frac1{E^*_j}~.
$$
The modified Young modulus is determined from the Young modulus $E$ and
the Poisson ratio $\nu$ of the material that the grains are made of:
$$
E^*=\frac{E}{1-\nu^2}~.
$$
Now let us determine a closed expression for the prefactor
$\epsilon_{ij}$ of the interaction as we have used it throughout the
paper. Differentiating the interaction in equation~(\ref{hertzpot}) gives
\begin{equation}
\label{hertzforce}
f_{ij}=\frac52\frac{\epsilon_{ij}}{d_{ij}}\left(1-\frac{r_{ij}}{d_{ij}}\right)^{3/2}~,
\end{equation}
with $d_{ij}=R_i+R_j$.
Equating the equations~(\ref{HertzJohnson}) and (\ref{hertzforce}) yields
$$
\epsilon_{ij}=\frac8{15}\sqrt{R_{ij}}E^*_{ij}d_{ij}^{5/2}~.
$$
Our unit of pressure is $E^*_i$ (and all grains are made from the same
material), so that in our reduced units $E^*_{ij}=\frac12$ and
\begin{equation}
\label{epsilonhertz}
\epsilon_{ij}=\frac4{15}\sqrt{R_{ij}}d_{ij}^{5/2}~.
\end{equation}
With the radii of the particles drawn from a flat distribution between
$0.4\le R_i\le 0.6$, the value of $\epsilon_{ij}$ can vary between 0.068 and
0.230. To check how large the effect of this varying $\epsilon$ is, we
have performed some simulations using a constant $\epsilon=\frac2{15}$, the
value it has for $R=0.5$, and did not observe significant deviations.
This justifies comparing our results to other simulations done at
constant $\epsilon$~\cite{epitome}.

\section{2D elasticity and point response}
\label{app.Elas}
The two-dimensional version of linear elasticity has the same form for
Hooke's law as the well-known 3D version~\cite{lanlif7,benny}
$$
\sigma_{\alpha\beta}=2\mu u_{\alpha\beta}+\lambda u_{\xi\xi}\delta_{\alpha\beta}~,
$$
where $u_{\alpha\beta}$ is the strain tensor, $\delta_{\alpha\beta}$ is the Kronecker
delta, and we use the summation convention. $\lambda$ and $\mu$ are known
as the Lam\'e coefficients. All dependence on dimensionality basically
arises from the $\delta_{\alpha\beta}$, through the fact that the trace of
$\sigma_{\alpha\beta}$ reads
$$
\sigma_{\alpha\alpha}=(2\mu+d\lambda)u_{\alpha\alpha}~.
$$
The shear modulus $G$ is just the same as the Lam\'e coefficient
$\mu$, independent of $d$. The definition of the bulk modulus $K$ is
the resistance to change of ``volume'',
$$
K=-\frac{p}{V/V_0}~.
$$
In an isotropic material the stress tensor corresponding to a uniform compression
is $\sigma_{\alpha\beta}=p\delta_{\alpha\beta}$. Its trace is therefore
$\sigma_{\alpha\alpha}=pd$. The trace
of the strain tensor is $u_{\alpha\alpha}=V/V_0$, so that in $d=2$, we now have for
the moduli
\begin{eqnarray}
G&=&\mu~.\\
K&=&\frac{\sigma_{\alpha\alpha}}{du_{\alpha\alpha}}=\lambda+\frac2d\mu=\lambda+\mu~.
\end{eqnarray}
The Navier-Cauchy equation, which follows from inserting Hooke's law and
the definition of the strain tensor into the equation of mechanical
equilibrium $\partial_\beta\sigma_{\alpha\beta}=-f_\alpha^\mathrm{ext}$, reads (in
arbitrary dimension),
$$
\mu\Delta\vc{u}+(\lambda+\mu)\nabla(\nabla\cdot\vc{u})=-\vc{f}^\mathrm{ext}~,
$$
where $\Delta$ denotes the Laplacian operator. In $d=2$ the coefficients
happen to become simply
\begin{equation}
\label{navcauapp}
G\Delta\vc{u}+K\nabla(\nabla\cdot\vc{u})=-\vc{f}^\mathrm{ext}~,
\end{equation}
which is equation~(\ref{navcau}).

\subsection{Response to a point force}
For a rectangular packing with periodic $x$-boundaries and hard walls
on top and bottom, the Navier-Cauchy equation (\ref{navcauapp}) can be
solved in terms of a Fourier series. For a point force in the middle,
taken to be the origin of the coordinate system, we expand:
\begin{eqnarray}
u_x&=&\sum_{lm}A_{lm}\sin\left(\frac{2\pi
lx}{L_x}\right)\sin\left(\frac{\pi my}{L_y}\right)\\
u_y&=&\sum_{lm}B_{lm}\cos\left(\frac{2\pi
lx}{L_x}\right)\cos\left(\frac{\pi my}{L_y}\right)\\
f_x&=&0\\
f_y&=&\sum_{lm}F_{lm}\cos\left(\frac{2\pi
lx}{L_x}\right)\cos\left(\frac{\pi my}{L_y}\right),
\end{eqnarray}
where
\begin{equation}
\label{ffourier}
F_{lm}=\left\{
\begin{array}{ll}
\frac{2}{L_xL_y}&m~\mathrm{odd}\\
0&m~\mathrm{even}
\end{array}
\right.,
\end{equation}
so $f_y$ represents a $\delta$-force in the origin of unit weight. The
solution is to be compared to granular stress fields which have been
coarse grained using a Gaussian. Using the same Gaussian instead of a
$\delta$-function for $f_y$ here leads to a better fit and to improved
convergence of the resulting Fourier series for the stress tensor. Since
convolution in real space is just multiplication in Fourier space this
amounts to multiplying $F_{lm}$ by
$$
\frac{2}{L_xL_y}\exp\left[-\pi^2\sigma^2\left(\frac{2l^2}{L_x^2}-\frac{m^2}{2L_y^2}\right)\right]
$$

Inserting the expansions into equation~(\ref{navcauapp}) we can solve for
$A_{lm}$ and $B_{lm}$. Solutions for the strain tensor and stress
tensor are then obtained by taking the appropriate linear combinations
of derivatives of $u_x$ and $u_y$. The resulting stress tensor only
depends on the Poisson ratio $\nu=(K-G)/(K+G)$. A separate fitting
procedure for the displacement field then gives $K+G$, using
equation~(\ref{exactHy}) and the determination of the moduli is complete.

\section{Floppy Modes}\label{app.floppy} Suppose we are at point J, so $\delta =0$,
$z = 4$, and the number of contacts equals $2N$, which is equal to
the number of independent displacement components $u_{i,\alpha}$.
When one contact is removed, we can in principle write down a
displacement field for which all $u_{\parallel,ij}=0$. This
displacement field has energy zero: it is a \emph{floppy mode}
\cite{alexander,wyartEPL}. This is not merely an artifact of the
expansion in $\uparl$ and $\uperp$: the counting of variables and
equations in principle allows to write down a displacement field
for which the unexpanded $\Delta r_{ij}$ from Eq.~(\ref{Deltarij})
are zero and hence the change in energy is strictly zero.
For a floppy mode distortion, $\Delta E$ is identically zero to
all orders in the distortion, even if the initial configuration
did not obey force balance. Let us check this up to
$\mathcal{O}(\uperp^4)$ in the expansion of $\Delta E$. To see
this, we include the term linear in $\uparl$, which we left out of
the original equation~(\ref{DeltaE}) because it vanishes when
summed over all contacts:
$$
\Delta E=\sum_{\langle i,j\rangle}\left[ -f_{ij}u_{\parallel, ij}+
\frac12k_{ij}\left(u_{\parallel,ij}^2-\frac{f_{ij}}{k_{ij}r_{ij}}u_{\perp,ij}^2\right)\right]~.
$$
For a floppy mode, $\Delta r_{ij}=0$, and this implies that
$\uparlij=-\uperpij^2/(2r_{ij})+\mathcal{O}(\uperp^4)$. Hence, the
first and third term in the expansion of $\Delta E$ cancel without
having to sum over all contacts and we are left with $\Delta
E=\mathcal{O}(\uperp^4)$. This even holds  if the initial system
would not satisfy force balance~\footnote{This holds in general
for floppy modes in systems with radial interactions between the
particles.}.

\section{Finite size effects} \label{subsec.finite}
If the
system is sheared or compressed, local relative displacements
diverge upon approaching the jamming point, according to the analysis in
section~\ref{subsec.udiverge}.  However, the relative displacement
between a given particle $i$ and its image in the neighboring unit
cell of the periodic packing is given by the imposed boundary
condition --- hence there is coupling between the magnitude of the
local relative displacements and the system size. In addition, we
have extracted the length scale $\ell^*$ from the elastic behavior
of our system, and should consider what happens when this length
scale becomes of the order of the system size. We tentatively
suggest the following picture for the implications of this
coupling.

Let us focus on shear deformations because the scalings are
cleanest in that case. The $x$-component of the relative
displacement between a particle at $(x_i,y_i)$ and its periodic
image at $(x_i,y_i+L)$ can be written as a sum along a path from
the particle to its image:
\begin{equation}
\label{finitesum} \sum_\mathrm{path} u_{ij,x}=\epsilon L~,
\end{equation}
where $\epsilon$ is the applied shear strain. If the system is
small compared to the length scale $\ell^*$, we do not expect to
see any elastic behavior, and the displacement field is dominated
by the local packing disorder. In this case, we might just as well
view the displacements in the $x$-direction as we go along the
path as a random walk. This walk consists of approximately $L$
steps of size roughly $\uperp$, so that
\begin{equation}
\label{finitesumwalk} \sum_\mathrm{path}
u_{ij,x}\sim\uperp\sqrt{L}~.
\end{equation}
Equating (\ref{finitesum}) and (\ref{finitesumwalk}) yields
\begin{equation}
\label{uperpfinitesmall}
\uperp\sim\epsilon\sqrt{L}\qquad\mathrm{for~}L\ll\ell^*~.
\end{equation}
On the other hand, large systems can be treated as a continuum
down to the scale $\ell^*$, so if we want to apply this model to a
large system, we have to consider the ``random walk'' in a
subsystem of size $\ell^*$, and assume the globally applied shear
deformation scales down affinely to that scale. Therefore the
boundary condition becomes
\begin{equation}
\label{finite2sum} \sum_\mathrm{subsystem}
u_{ij,x}=\epsilon\ell^*~,
\end{equation}
and the random walk in the subsystem gives
\begin{equation}
\label{finite2sumwalk} \sum_\mathrm{subsystem}
u_{ij,x}\sim\uperp\sqrt{\ell^*}~,
\end{equation}
so that finally
\begin{equation}
\label{uperpfinitelarge}
\uperp\sim\epsilon\sqrt{\ell^*}\qquad\mathrm{for~}L\gg\ell^*~.
\end{equation}
\begin{figure}[!bt]
\includegraphics[width=8.4cm]{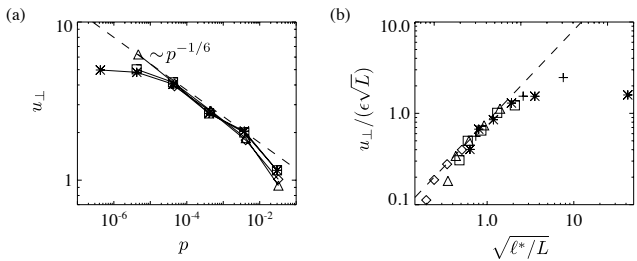}
\caption{(a) Scaling of typical $\uperp$ under shear with pressure
$p$. The dashed line indicates $\uperp\sim p^{-1/6}$, to which all
large packings stay close, except for the bottom-right points
which are very far from point J ($p=3\cdot10^{-2}$). The system
size decreases from $N=10\,000$ (diamonds), $N=1000$ (triangles),
$N=300$ (squares) to $N=100$ (stars). The curve seems to level off
at small $p$ for the small systems, as predicted in
equation~(\ref{uperpfinitesmall}). (b)~The same data, plotted
according to the finite size scaling prediction of
equation~(\ref{fssuperp}). The plus signs indicate data taken from
the linear response to shear of bidisperse packings generated
using Conjugate Gradient minimization (courtesy N. Xu), which
allows to probe much lower pressures. The two rightmost points in
this graph represent series of packings which are essentially
isostatic, and hence have a very large and inaccurate $\ell^*$.
The dashed line represents the predicted behavior for small
$\ell^*/L$.} \label{fig:fssuperp}
\end{figure}%
Hence the argument reproduces our result for the scaling of
$\uperp$ under shear, equation~(\ref{uperpshearpred}), because
$\ell^*\sim \delta^{-1/2}$. Furthermore, it predicts that $\uperp$
should saturate around the value it has when $L=\ell^*$ when
approaching the jamming transition in a finite system. The two
limits can be connected by a finite size scaling function
\begin{equation}
\label{fssuperp} \uperp\approx\epsilon\sqrt{L
}f\left(\sqrt{\frac{\ell^*}{L}}\right)~,
\end{equation}
where $f(x)\to\mathrm{const}$ as $x\gg1$ and $f(x)\sim x$ as
$x\ll1$. We do not have enough data on systems where $L\ll\ell^*$
to prove this scaling prediction with firm confidence, but the
data we have is at least consistent with it. In
Fig.~\ref{fig:fssuperp}, we show the typical values of $\uperp$ as
a function of $p$ for various system sizes, taking
$\ell^*=6/\Delta z$ in accord with (\ref{ellexpression}).
Figure~\ref{fig:fssuperp}b shows the scaling collapse
corresponding to equation~(\ref{fssuperp}). The two rightmost
points in this plot have $z\approx4$, so $\ell^*$ is very large
and very inaccurate, but nevertheless some levelling off of
the curve can be observed. More data using packings between 100
and 1000 particles at $4.005\le z\le4.02$ could shed more light on
the scaling behavior.


\begin{thebibliography}{53}
\expandafter\ifx\csname natexlab\endcsname\relax\def\natexlab#1{#1}\fi
\expandafter\ifx\csname bibnamefont\endcsname\relax
  \def\bibnamefont#1{#1}\fi
\expandafter\ifx\csname bibfnamefont\endcsname\relax
  \def\bibfnamefont#1{#1}\fi
\expandafter\ifx\csname citenamefont\endcsname\relax
  \def\citenamefont#1{#1}\fi
\expandafter\ifx\csname url\endcsname\relax
  \def\url#1{\texttt{#1}}\fi
\expandafter\ifx\csname urlprefix\endcsname\relax\def\urlprefix{URL }\fi
\providecommand{\bibinfo}[2]{#2}
\providecommand{\eprint}[2][]{\url{#2}}

\bibitem[{\citenamefont{Liu and Nagel}(1998)}]{jamnature}
\bibinfo{author}{\bibfnamefont{A.~J.} \bibnamefont{Liu}} \bibnamefont{and}
  \bibinfo{author}{\bibfnamefont{S.~R.} \bibnamefont{Nagel}},
  \bibinfo{journal}{Nature} \textbf{\bibinfo{volume}{396}}, \bibinfo{pages}{21}
  (\bibinfo{year}{1998}).

\bibitem[{\citenamefont{Trappe et~al.}(2001)\citenamefont{Trappe, Prasad,
  Cipelletti, Segre, and Weitz}}]{trappe}
\bibinfo{author}{\bibfnamefont{V.}~\bibnamefont{Trappe}},
  \bibinfo{author}{\bibfnamefont{V.}~\bibnamefont{Prasad}},
  \bibinfo{author}{\bibfnamefont{L.}~\bibnamefont{Cipelletti}},
  \bibinfo{author}{\bibfnamefont{P.~N.} \bibnamefont{Segre}}, \bibnamefont{and}
  \bibinfo{author}{\bibfnamefont{D.~A.} \bibnamefont{Weitz}},
  \bibinfo{journal}{Nature} \textbf{\bibinfo{volume}{411}},
  \bibinfo{pages}{772} (\bibinfo{year}{2001}).

\bibitem[{\citenamefont{O'Hern et~al.}(2003)\citenamefont{O'Hern, Silbert, Liu,
  and Nagel}}]{epitome}
\bibinfo{author}{\bibfnamefont{C.~S.} \bibnamefont{O'Hern}},
  \bibinfo{author}{\bibfnamefont{L.~E.} \bibnamefont{Silbert}},
  \bibinfo{author}{\bibfnamefont{A.~J.} \bibnamefont{Liu}}, \bibnamefont{and}
  \bibinfo{author}{\bibfnamefont{S.~R.} \bibnamefont{Nagel}},
  \bibinfo{journal}{Phys.\ Rev.\ E} \textbf{\bibinfo{volume}{68}},
  \bibinfo{pages}{011306} (\bibinfo{year}{2003}).

\bibitem[{\citenamefont{Silbert et~al.}(2005)\citenamefont{Silbert, Liu, and
  Nagel}}]{silbertPRL05}
\bibinfo{author}{\bibfnamefont{L.~E.} \bibnamefont{Silbert}},
  \bibinfo{author}{\bibfnamefont{A.~J.} \bibnamefont{Liu}}, \bibnamefont{and}
  \bibinfo{author}{\bibfnamefont{S.~R.} \bibnamefont{Nagel}},
  \bibinfo{journal}{Phys.\ Rev.\ Lett.} \textbf{\bibinfo{volume}{95}},
  \bibinfo{pages}{098301} (\bibinfo{year}{2005}).

\bibitem[{\citenamefont{Silbert et~al.}(2006)\citenamefont{Silbert, Liu, and
  Nagel}}]{silbertPRE06}
\bibinfo{author}{\bibfnamefont{L.~E.} \bibnamefont{Silbert}},
  \bibinfo{author}{\bibfnamefont{A.~J.} \bibnamefont{Liu}}, \bibnamefont{and}
  \bibinfo{author}{\bibfnamefont{S.~R.} \bibnamefont{Nagel}},
  \bibinfo{journal}{Phys.\ Rev.\ E} \textbf{\bibinfo{volume}{73}},
  \bibinfo{pages}{041304} (\bibinfo{year}{2006}).

\bibitem[{\citenamefont{Wyart et~al.}(2005{\natexlab{a}})\citenamefont{Wyart,
  Nagel, and Witten}}]{wyartEPL}
\bibinfo{author}{\bibfnamefont{M.}~\bibnamefont{Wyart}},
  \bibinfo{author}{\bibfnamefont{S.~R.} \bibnamefont{Nagel}}, \bibnamefont{and}
  \bibinfo{author}{\bibfnamefont{T.~A.} \bibnamefont{Witten}},
  \bibinfo{journal}{Europhys.\ Lett.} \textbf{\bibinfo{volume}{72}},
  \bibinfo{pages}{486} (\bibinfo{year}{2005}{\natexlab{a}}).

\bibitem[{\citenamefont{Wyart et~al.}(2005{\natexlab{b}})\citenamefont{Wyart,
  Silbert, Nagel, and Witten}}]{wyartPRE}
\bibinfo{author}{\bibfnamefont{M.}~\bibnamefont{Wyart}},
  \bibinfo{author}{\bibfnamefont{L.~E.} \bibnamefont{Silbert}},
  \bibinfo{author}{\bibfnamefont{S.~R.} \bibnamefont{Nagel}}, \bibnamefont{and}
  \bibinfo{author}{\bibfnamefont{T.~A.} \bibnamefont{Witten}},
  \bibinfo{journal}{Phys.\ Rev.\ E} \textbf{\bibinfo{volume}{72}},
  \bibinfo{eid}{051306} (\bibinfo{year}{2005}{\natexlab{b}}).

\bibitem[{\citenamefont{Ellenbroek et~al.}(2006)\citenamefont{Ellenbroek,
  Somfai, van Hecke, and van Saarloos}}]{respprl}
\bibinfo{author}{\bibfnamefont{W.~G.} \bibnamefont{Ellenbroek}},
  \bibinfo{author}{\bibfnamefont{E.}~\bibnamefont{Somfai}},
  \bibinfo{author}{\bibfnamefont{M.}~\bibnamefont{van Hecke}},
  \bibnamefont{and} \bibinfo{author}{\bibfnamefont{W.}~\bibnamefont{van
  Saarloos}}, \bibinfo{journal}{Phys.\ Rev.\ Lett.}
  \textbf{\bibinfo{volume}{97}}, \bibinfo{eid}{258001} (\bibinfo{year}{2006}).

\bibitem[{\citenamefont{Bolton and Weaire}(1990)}]{bolton}
\bibinfo{author}{\bibfnamefont{F.}~\bibnamefont{Bolton}} \bibnamefont{and}
  \bibinfo{author}{\bibfnamefont{D.}~\bibnamefont{Weaire}},
  \bibinfo{journal}{Phys. Rev. Lett.} \textbf{\bibinfo{volume}{65}},
  \bibinfo{pages}{3449} (\bibinfo{year}{1990}).

\bibitem[{\citenamefont{Durian}(1997)}]{durianbubble}
\bibinfo{author}{\bibfnamefont{D.~J.} \bibnamefont{Durian}},
  \bibinfo{journal}{Phys.\ Rev.\ E} \textbf{\bibinfo{volume}{55}},
  \bibinfo{pages}{1739} (\bibinfo{year}{1997}).

\bibitem[{\citenamefont{Weaire and Hutzler}(2001)}]{weairebook}
\bibinfo{author}{\bibfnamefont{D.}~\bibnamefont{Weaire}} \bibnamefont{and}
  \bibinfo{author}{\bibfnamefont{S.}~\bibnamefont{Hutzler}},
  \emph{\bibinfo{title}{The Physics of Foams}} (\bibinfo{publisher}{Oxford
  University Press}, \bibinfo{address}{Oxford}, \bibinfo{year}{2001}).

\bibitem[{\citenamefont{Makse et~al.}(1999)\citenamefont{Makse, Gland, Johnson,
  and Schwartz}}]{maksePRL}
\bibinfo{author}{\bibfnamefont{H.~A.} \bibnamefont{Makse}},
  \bibinfo{author}{\bibfnamefont{N.}~\bibnamefont{Gland}},
  \bibinfo{author}{\bibfnamefont{D.~L.} \bibnamefont{Johnson}},
  \bibnamefont{and} \bibinfo{author}{\bibfnamefont{L.~M.}
  \bibnamefont{Schwartz}}, \bibinfo{journal}{Phys.\ Rev.\ Lett.}
  \textbf{\bibinfo{volume}{83}}, \bibinfo{pages}{5070} (\bibinfo{year}{1999}).

\bibitem[{\citenamefont{Somfai et~al.}(2007)\citenamefont{Somfai, van Hecke,
  Ellenbroek, Shundyak, and van Saarloos}}]{ellakjam}
\bibinfo{author}{\bibfnamefont{E.}~\bibnamefont{Somfai}},
  \bibinfo{author}{\bibfnamefont{M.}~\bibnamefont{van Hecke}},
  \bibinfo{author}{\bibfnamefont{W.~G.} \bibnamefont{Ellenbroek}},
  \bibinfo{author}{\bibfnamefont{K.}~\bibnamefont{Shundyak}}, \bibnamefont{and}
  \bibinfo{author}{\bibfnamefont{W.}~\bibnamefont{van Saarloos}},
  \bibinfo{journal}{Phys.\ Rev.\ E} \textbf{\bibinfo{volume}{75}},
  \bibinfo{eid}{020301(R)} (\bibinfo{year}{2007}).

\bibitem[{\citenamefont{Shundyak et~al.}(2007)\citenamefont{Shundyak, van
  Hecke, and van Saarloos}}]{kostyaPRE07}
\bibinfo{author}{\bibfnamefont{K.}~\bibnamefont{Shundyak}},
  \bibinfo{author}{\bibfnamefont{M.}~\bibnamefont{van Hecke}},
  \bibnamefont{and} \bibinfo{author}{\bibfnamefont{W.}~\bibnamefont{van
  Saarloos}}, \bibinfo{journal}{Phys.\ Rev.\ E} \textbf{\bibinfo{volume}{75}},
  \bibinfo{pages}{010301(R)} (\bibinfo{year}{2007}).

\bibitem[{\citenamefont{Donev et~al.}(2004)\citenamefont{Donev, Cisse, Sachs,
  Variano, Stillinger, Connelly, Torquato, and Chaikin}}]{donev04}
\bibinfo{author}{\bibfnamefont{A.}~\bibnamefont{Donev}},
  \bibinfo{author}{\bibfnamefont{I.}~\bibnamefont{Cisse}},
  \bibinfo{author}{\bibfnamefont{D.}~\bibnamefont{Sachs}},
  \bibinfo{author}{\bibfnamefont{E.~A.} \bibnamefont{Variano}},
  \bibinfo{author}{\bibfnamefont{F.~H.} \bibnamefont{Stillinger}},
  \bibinfo{author}{\bibfnamefont{R.}~\bibnamefont{Connelly}},
  \bibinfo{author}{\bibfnamefont{S.}~\bibnamefont{Torquato}}, \bibnamefont{and}
  \bibinfo{author}{\bibfnamefont{P.~M.} \bibnamefont{Chaikin}},
  \bibinfo{journal}{Science} \textbf{\bibinfo{volume}{303}},
  \bibinfo{pages}{990} (\bibinfo{year}{2004}).

\bibitem[{\citenamefont{Wouterse et~al.}(2007)\citenamefont{Wouterse, Williams,
  and Philipse}}]{wouterse07}
\bibinfo{author}{\bibfnamefont{A.}~\bibnamefont{Wouterse}},
  \bibinfo{author}{\bibfnamefont{S.~R.} \bibnamefont{Williams}},
  \bibnamefont{and} \bibinfo{author}{\bibfnamefont{A.~P.}
  \bibnamefont{Philipse}}, \bibinfo{journal}{J. Phys.: Condens.\ Matter}
  \textbf{\bibinfo{volume}{19}}, \bibinfo{pages}{406215}
  (\bibinfo{year}{2007}).

\bibitem[{\citenamefont{Zeravcic et~al.}(2009)\citenamefont{Zeravcic, Xu, Liu,
  Nagel, and van Saarloos}}]{zorana09}
\bibinfo{author}{\bibfnamefont{Z.}~\bibnamefont{Zeravcic}},
  \bibinfo{author}{\bibfnamefont{N.}~\bibnamefont{Xu}},
  \bibinfo{author}{\bibfnamefont{A.~J.} \bibnamefont{Liu}},
  \bibinfo{author}{\bibfnamefont{S.~R.} \bibnamefont{Nagel}}, \bibnamefont{and}
  \bibinfo{author}{\bibfnamefont{W.}~\bibnamefont{van Saarloos}},
  \bibinfo{journal}{EPL} \textbf{\bibinfo{volume}{87}}, \bibinfo{pages}{26001}
  (\bibinfo{year}{2009}).


\bibitem[{\citenamefont{Mailman et~al.}(2009)\citenamefont{Mailman, Schreck,
  O'Hern, and Chakraborty}}]{bulbul09}
\bibinfo{author}{\bibfnamefont{M.}~\bibnamefont{Mailman}},
  \bibinfo{author}{\bibfnamefont{C.~F.} \bibnamefont{Schreck}},
  \bibinfo{author}{\bibfnamefont{C.~S.} \bibnamefont{O'Hern}},
  \bibnamefont{and}
  \bibinfo{author}{\bibfnamefont{B.}~\bibnamefont{Chakraborty}},
  \bibinfo{journal}{Phys.\ Rev.\ Lett.} \textbf{\bibinfo{volume}{102}},
  \bibinfo{eid}{255501} (\bibinfo{year}{2009}).

\bibitem[{\citenamefont{Goldhirsch and Goldenberg}(2002)}]{goldcoarse}
\bibinfo{author}{\bibfnamefont{I.}~\bibnamefont{Goldhirsch}} \bibnamefont{and}
  \bibinfo{author}{\bibfnamefont{C.}~\bibnamefont{Goldenberg}},
  \bibinfo{journal}{Eur.\ Phys.\ J E} \textbf{\bibinfo{volume}{9}},
  \bibinfo{pages}{245} (\bibinfo{year}{2002}).

\bibitem[{\citenamefont{Goldenberg and Goldhirsch}(2002)}]{goldPRL02}
\bibinfo{author}{\bibfnamefont{C.}~\bibnamefont{Goldenberg}} \bibnamefont{and}
  \bibinfo{author}{\bibfnamefont{I.}~\bibnamefont{Goldhirsch}},
  \bibinfo{journal}{Phys.\ Rev.\ Lett.} \textbf{\bibinfo{volume}{89}},
  \bibinfo{pages}{084302} (\bibinfo{year}{2002}).

\bibitem[{\citenamefont{Goldenberg and Goldhirsch}(2005)}]{goldnature}
\bibinfo{author}{\bibfnamefont{C.}~\bibnamefont{Goldenberg}} \bibnamefont{and}
  \bibinfo{author}{\bibfnamefont{I.}~\bibnamefont{Goldhirsch}},
  \bibinfo{journal}{Nature} \textbf{\bibinfo{volume}{435}},
  \bibinfo{pages}{188} (\bibinfo{year}{2005}).

\bibitem[{\citenamefont{Goldenberg et~al.}(2006)\citenamefont{Goldenberg,
  Atman, Claudin, Combe, and Goldhirsch}}]{goldPRL06}
\bibinfo{author}{\bibfnamefont{C.}~\bibnamefont{Goldenberg}},
  \bibinfo{author}{\bibfnamefont{A.~P.~F.} \bibnamefont{Atman}},
  \bibinfo{author}{\bibfnamefont{P.}~\bibnamefont{Claudin}},
  \bibinfo{author}{\bibfnamefont{G.}~\bibnamefont{Combe}}, \bibnamefont{and}
  \bibinfo{author}{\bibfnamefont{I.}~\bibnamefont{Goldhirsch}},
  \bibinfo{journal}{Phys.\ Rev.\ Lett.} \textbf{\bibinfo{volume}{96}},
  \bibinfo{eid}{168001} (\bibinfo{year}{2006}).

\bibitem[{\citenamefont{Vanel et~al.}(1999)\citenamefont{Vanel, Howell, Clark,
  Behringer, and Cl\'ement}}]{vanel}
\bibinfo{author}{\bibfnamefont{L.}~\bibnamefont{Vanel}},
  \bibinfo{author}{\bibfnamefont{D.}~\bibnamefont{Howell}},
  \bibinfo{author}{\bibfnamefont{D.}~\bibnamefont{Clark}},
  \bibinfo{author}{\bibfnamefont{R.~P.} \bibnamefont{Behringer}},
  \bibnamefont{and}
  \bibinfo{author}{\bibfnamefont{E.}~\bibnamefont{Cl\'ement}},
  \bibinfo{journal}{Phys.\ Rev.\ E} \textbf{\bibinfo{volume}{60}},
  \bibinfo{pages}{R5040} (\bibinfo{year}{1999}).

\bibitem[{\citenamefont{Serero et~al.}(2001)\citenamefont{Serero, Reydellet,
  Claudin, Cl\'ement, and Levine}}]{serero01}
\bibinfo{author}{\bibfnamefont{D.}~\bibnamefont{Serero}},
  \bibinfo{author}{\bibfnamefont{G.}~\bibnamefont{Reydellet}},
  \bibinfo{author}{\bibfnamefont{P.}~\bibnamefont{Claudin}},
  \bibinfo{author}{\bibfnamefont{E.}~\bibnamefont{Cl\'ement}},
  \bibnamefont{and} \bibinfo{author}{\bibfnamefont{D.}~\bibnamefont{Levine}},
  \bibinfo{journal}{Eur.\ Phys.\ J E} \textbf{\bibinfo{volume}{6}},
  \bibinfo{pages}{169} (\bibinfo{year}{2001}).

\bibitem[{\citenamefont{Geng et~al.}(2001)\citenamefont{Geng, Howell, Longhi,
  Behringer, Reydellet, Vanel, Cl\'ement, and Luding}}]{gengPRL01}
\bibinfo{author}{\bibfnamefont{J.}~\bibnamefont{Geng}},
  \bibinfo{author}{\bibfnamefont{D.}~\bibnamefont{Howell}},
  \bibinfo{author}{\bibfnamefont{E.}~\bibnamefont{Longhi}},
  \bibinfo{author}{\bibfnamefont{R.~P.} \bibnamefont{Behringer}},
  \bibinfo{author}{\bibfnamefont{G.}~\bibnamefont{Reydellet}},
  \bibinfo{author}{\bibfnamefont{L.}~\bibnamefont{Vanel}},
  \bibinfo{author}{\bibfnamefont{E.}~\bibnamefont{Cl\'ement}},
  \bibnamefont{and} \bibinfo{author}{\bibfnamefont{S.}~\bibnamefont{Luding}},
  \bibinfo{journal}{Phys.\ Rev.\ Lett.} \textbf{\bibinfo{volume}{87}},
  \bibinfo{pages}{035506} (\bibinfo{year}{2001}).

\bibitem[{\citenamefont{Reydellet and Cl\'ement}(2001)}]{reydellet01}
\bibinfo{author}{\bibfnamefont{G.}~\bibnamefont{Reydellet}} \bibnamefont{and}
  \bibinfo{author}{\bibfnamefont{E.}~\bibnamefont{Cl\'ement}},
  \bibinfo{journal}{Phys.\ Rev.\ Lett.} \textbf{\bibinfo{volume}{86}},
  \bibinfo{pages}{3308} (\bibinfo{year}{2001}).

\bibitem[{\citenamefont{Otto et~al.}(2003)\citenamefont{Otto, Bouchaud,
  Claudin, and Socolar}}]{ottoPRE03}
\bibinfo{author}{\bibfnamefont{M.}~\bibnamefont{Otto}},
  \bibinfo{author}{\bibfnamefont{J.-P.} \bibnamefont{Bouchaud}},
  \bibinfo{author}{\bibfnamefont{P.}~\bibnamefont{Claudin}}, \bibnamefont{and}
  \bibinfo{author}{\bibfnamefont{J.~E.~S.} \bibnamefont{Socolar}},
  \bibinfo{journal}{Phys.\ Rev.\ E} \textbf{\bibinfo{volume}{67}},
  \bibinfo{pages}{031302} (\bibinfo{year}{2003}).

\bibitem[{\citenamefont{Atman et~al.}(2005)\citenamefont{Atman, Brunet, Geng,
  Reydellet, Claudin, Behringer, and Cl\'ement}}]{atmanbrunet}
\bibinfo{author}{\bibfnamefont{A.~P.~F.} \bibnamefont{Atman}},
  \bibinfo{author}{\bibfnamefont{P.}~\bibnamefont{Brunet}},
  \bibinfo{author}{\bibfnamefont{J.}~\bibnamefont{Geng}},
  \bibinfo{author}{\bibfnamefont{G.}~\bibnamefont{Reydellet}},
  \bibinfo{author}{\bibfnamefont{P.}~\bibnamefont{Claudin}},
  \bibinfo{author}{\bibfnamefont{R.~P.} \bibnamefont{Behringer}},
  \bibnamefont{and}
  \bibinfo{author}{\bibfnamefont{E.}~\bibnamefont{Cl\'ement}},
  \bibinfo{journal}{Eur.\ Phys.\ J E} \textbf{\bibinfo{volume}{17}},
  \bibinfo{pages}{93} (\bibinfo{year}{2005}).

\bibitem[{\citenamefont{Ostojic and Panja}(2006)}]{srdjanPRL06}
\bibinfo{author}{\bibfnamefont{S.}~\bibnamefont{Ostojic}} \bibnamefont{and}
  \bibinfo{author}{\bibfnamefont{D.}~\bibnamefont{Panja}},
  \bibinfo{journal}{Phys.\ Rev.\ Lett.} \textbf{\bibinfo{volume}{97}},
  \bibinfo{pages}{208001} (\bibinfo{year}{2006}).

\bibitem[{\citenamefont{Makse et~al.}(2004)\citenamefont{Makse, Gland, Johnson,
  and Schwartz}}]{maksePRE}
\bibinfo{author}{\bibfnamefont{H.~A.} \bibnamefont{Makse}},
  \bibinfo{author}{\bibfnamefont{N.}~\bibnamefont{Gland}},
  \bibinfo{author}{\bibfnamefont{D.~L.} \bibnamefont{Johnson}},
  \bibnamefont{and} \bibinfo{author}{\bibfnamefont{L.~M.}
  \bibnamefont{Schwartz}}, \bibinfo{journal}{Phys.\ Rev.\ E}
  \textbf{\bibinfo{volume}{70}}, \bibinfo{eid}{061302} (\bibinfo{year}{2004}).

\bibitem[{\citenamefont{Ellenbroek et~al.}(2009)\citenamefont{Ellenbroek,
  Zeravcic, van Saarloos, and van Hecke}}]{wouterEPL09}
\bibinfo{author}{\bibfnamefont{W.~G.} \bibnamefont{Ellenbroek}},
  \bibinfo{author}{\bibfnamefont{Z.}~\bibnamefont{Zeravcic}},
  \bibinfo{author}{\bibfnamefont{W.}~\bibnamefont{van Saarloos}},
  \bibnamefont{and} \bibinfo{author}{\bibfnamefont{M.}~\bibnamefont{van
  Hecke}}, \bibinfo{journal}{EPL} \textbf{\bibinfo{volume}{87}},
  \bibinfo{pages}{34004} (\bibinfo{year}{2009}).


\bibitem[{\citenamefont{Somfai et~al.}(2005)\citenamefont{Somfai, Roux,
  Snoeijer, van Hecke, and van Saarloos}}]{ellakwave}
\bibinfo{author}{\bibfnamefont{E.}~\bibnamefont{Somfai}},
  \bibinfo{author}{\bibfnamefont{J.-N.} \bibnamefont{Roux}},
  \bibinfo{author}{\bibfnamefont{J.~H.} \bibnamefont{Snoeijer}},
  \bibinfo{author}{\bibfnamefont{M.}~\bibnamefont{van Hecke}},
  \bibnamefont{and} \bibinfo{author}{\bibfnamefont{W.}~\bibnamefont{van
  Saarloos}}, \bibinfo{journal}{Phys.\ Rev.\ E} \textbf{\bibinfo{volume}{72}},
  \bibinfo{eid}{021301} (\bibinfo{year}{2005}).

\bibitem[{\citenamefont{Leonforte et~al.}(2004)\citenamefont{Leonforte, Tanguy,
  Wittmer, and Barrat}}]{leonforte}
\bibinfo{author}{\bibfnamefont{F.}~\bibnamefont{Leonforte}},
  \bibinfo{author}{\bibfnamefont{A.}~\bibnamefont{Tanguy}},
  \bibinfo{author}{\bibfnamefont{J.~P.} \bibnamefont{Wittmer}},
  \bibnamefont{and}
  \bibinfo{author}{\bibfnamefont{J.-L.} \bibnamefont{Barrat}},
  \bibinfo{journal}{Phys.\ Rev.\ B} \textbf{\bibinfo{volume}{70}},
  \bibinfo{pages}{014203} (\bibinfo{year}{2004}).

\bibitem[{\citenamefont{Ellenbroek et~al.}(2005)\citenamefont{Ellenbroek,
  Somfai, van Saarloos, and van Hecke}}]{wouterpowders}
\bibinfo{author}{\bibfnamefont{W.~G.} \bibnamefont{Ellenbroek}},
  \bibinfo{author}{\bibfnamefont{E.}~\bibnamefont{Somfai}},
  \bibinfo{author}{\bibfnamefont{W.}~\bibnamefont{van Saarloos}},
  \bibnamefont{and} \bibinfo{author}{\bibfnamefont{M.}~\bibnamefont{van
  Hecke}}, in \emph{\bibinfo{booktitle}{Powders and Grains 2005}}, edited by
  \bibinfo{editor}{\bibfnamefont{R.}~\bibnamefont{Garc\'ia-Rojo}},
  \bibinfo{editor}{\bibfnamefont{H.~J.} \bibnamefont{Herrmann}},
  \bibnamefont{and} \bibinfo{editor}{\bibfnamefont{S.}~\bibnamefont{McNamara}}
  (\bibinfo{publisher}{A. A. Balkema Publishers}, \bibinfo{address}{Rotterdam,
  The Netherlands}, \bibinfo{year}{2005}), pp. \bibinfo{pages}{377--380}.

\bibitem[{\citenamefont{Tanguy et~al.}(2002)\citenamefont{Tanguy, Wittmer,
  Leonforte, and Barrat}}]{tanguy}
\bibinfo{author}{\bibfnamefont{A.}~\bibnamefont{Tanguy}},
  \bibinfo{author}{\bibfnamefont{J.~P.} \bibnamefont{Wittmer}},
  \bibinfo{author}{\bibfnamefont{F.}~\bibnamefont{Leonforte}},
  \bibnamefont{and} \bibinfo{author}{\bibfnamefont{J.-L.}
  \bibnamefont{Barrat}}, \bibinfo{journal}{Phys.\ Rev.\ B}
  \textbf{\bibinfo{volume}{66}}, \bibinfo{pages}{174205}
  (\bibinfo{year}{2002}).

\bibitem[{\citenamefont{Johnson}(1985)}]{contact}
\bibinfo{author}{\bibfnamefont{K.~L.} \bibnamefont{Johnson}},
  \emph{\bibinfo{title}{Contact Mechanics}} (\bibinfo{publisher}{Cambridge
  University Press}, \bibinfo{address}{Cambridge, U.K.}, \bibinfo{year}{1985}).

\bibitem[{\citenamefont{Alexander}(1998)}]{alexander}
\bibinfo{author}{\bibfnamefont{S.}~\bibnamefont{Alexander}},
  \bibinfo{journal}{Physics Reports} \textbf{\bibinfo{volume}{296}},
  \bibinfo{pages}{65} (\bibinfo{year}{1998}).

\bibitem[{\citenamefont{Barrett et~al.}(1994)\citenamefont{Barrett, Berry,
  Chan, Demmel, Donato, Dongarra, Eijkhout, Pozo, Romine, and der
  Vorst}}]{templates}
\bibinfo{author}{\bibfnamefont{R.}~\bibnamefont{Barrett}},
  \bibinfo{author}{\bibfnamefont{M.}~\bibnamefont{Berry}},
  \bibinfo{author}{\bibfnamefont{T.~F.} \bibnamefont{Chan}},
  \bibinfo{author}{\bibfnamefont{J.}~\bibnamefont{Demmel}},
  \bibinfo{author}{\bibfnamefont{J.}~\bibnamefont{Donato}},
  \bibinfo{author}{\bibfnamefont{J.}~\bibnamefont{Dongarra}},
  \bibinfo{author}{\bibfnamefont{V.}~\bibnamefont{Eijkhout}},
  \bibinfo{author}{\bibfnamefont{R.}~\bibnamefont{Pozo}},
  \bibinfo{author}{\bibfnamefont{C.}~\bibnamefont{Romine}}, \bibnamefont{and}
  \bibinfo{author}{\bibfnamefont{H.~V.} \bibnamefont{der Vorst}},
  \emph{\bibinfo{title}{Templates for the Solution of Linear Systems: Building
  Blocks for Iterative Methods}} (\bibinfo{publisher}{SIAM},
  \bibinfo{address}{Philadelphia, PA}, \bibinfo{year}{1994}).

\bibitem[{\citenamefont{Lees and Edwards}(1972)}]{leesedwards}
\bibinfo{author}{\bibfnamefont{A.~W.} \bibnamefont{Lees}} \bibnamefont{and}
  \bibinfo{author}{\bibfnamefont{S.~F.} \bibnamefont{Edwards}},
  \bibinfo{journal}{J. Phys.\ C} \textbf{\bibinfo{volume}{5}},
  \bibinfo{pages}{1921} (\bibinfo{year}{1972}).

\bibitem[{\citenamefont{Lautrup}(2005)}]{benny}
\bibinfo{author}{\bibfnamefont{B.}~\bibnamefont{Lautrup}},
  \emph{\bibinfo{title}{Physics of Continuous Matter}} (\bibinfo{publisher}{IoP
  Publishing}, \bibinfo{address}{Bristol, U.K.}, \bibinfo{year}{2005}).

\bibitem[{\citenamefont{Wyart}(2005)}]{wyartthesis}
\bibinfo{author}{\bibfnamefont{M.}~\bibnamefont{Wyart}},
  \bibinfo{journal}{Ann.\ Phys.\ Fr.} \textbf{\bibinfo{volume}{30, No.\ 3}},
  \bibinfo{pages}{1} (\bibinfo{year}{2005}).

\bibitem[{\citenamefont{Walton}(1987)}]{walton87}
\bibinfo{author}{\bibfnamefont{K.}~\bibnamefont{Walton}}, \bibinfo{journal}{J.
  Mech.\ Phys.\ Solids} \textbf{\bibinfo{volume}{35}}, \bibinfo{pages}{213}
  (\bibinfo{year}{1987}).

\bibitem[{\citenamefont{Wyart et~al.}(2008)\citenamefont{Wyart, Liang, Kabla,
  and Mahadevan}}]{wyart08}
\bibinfo{author}{\bibfnamefont{M.}~\bibnamefont{Wyart}},
  \bibinfo{author}{\bibfnamefont{H.}~\bibnamefont{Liang}},
  \bibinfo{author}{\bibfnamefont{A.}~\bibnamefont{Kabla}}, \bibnamefont{and}
  \bibinfo{author}{\bibfnamefont{L.}~\bibnamefont{Mahadevan}},
  \bibinfo{journal}{Phys.\ Rev.\ Lett.} \textbf{\bibinfo{volume}{101}},
  \bibinfo{eid}{215501} (\bibinfo{year}{2008}).

\bibitem[{\citenamefont{Radjai and Roux}(2002)}]{granulence}
\bibinfo{author}{\bibfnamefont{F.}~\bibnamefont{Radjai}} \bibnamefont{and}
  \bibinfo{author}{\bibfnamefont{S.}~\bibnamefont{Roux}},
  \bibinfo{journal}{Phys.\ Rev.\ Lett.} \textbf{\bibinfo{volume}{89}},
  \bibinfo{pages}{064302} (\bibinfo{year}{2002}).

\bibitem[{\citenamefont{Lema\^itre and Maloney}(2006)}]{lemaitre06}
\bibinfo{author}{\bibfnamefont{A.}~\bibnamefont{Lema\^itre}} \bibnamefont{and}
  \bibinfo{author}{\bibfnamefont{C.}~\bibnamefont{Maloney}},
  \bibinfo{journal}{J. Stat.\ Phys.} \textbf{\bibinfo{volume}{123}},
  \bibinfo{pages}{415} (\bibinfo{year}{2006}).

\bibitem[{\citenamefont{Tanguy et~al.}(2004)\citenamefont{Tanguy, Leonforte,
  Wittmer, and Barrat}}]{tanguy04}
\bibinfo{author}{\bibfnamefont{A.}~\bibnamefont{Tanguy}},
  \bibinfo{author}{\bibfnamefont{F.}~\bibnamefont{Leonforte}},
  \bibinfo{author}{\bibfnamefont{J.~P.} \bibnamefont{Wittmer}},
  \bibnamefont{and} \bibinfo{author}{\bibfnamefont{J.~L.}
  \bibnamefont{Barrat}}, \bibinfo{journal}{App.\ Surf.\ Sci.}
  \textbf{\bibinfo{volume}{226}}, \bibinfo{pages}{282} (\bibinfo{year}{2004}).

\bibitem[{\citenamefont{Maloney}(2006)}]{maloneyPRL06}
\bibinfo{author}{\bibfnamefont{C.~E.} \bibnamefont{Maloney}},
  \bibinfo{journal}{Phys.\ Rev.\ Lett.} \textbf{\bibinfo{volume}{97}},
  \bibinfo{eid}{035503} (\bibinfo{year}{2006}).

\bibitem[{\citenamefont{Blair}(2006)}]{blairprivate}
\bibinfo{author}{\bibfnamefont{D.~L.} \bibnamefont{Blair}},
  \bibinfo{howpublished}{private communication} (\bibinfo{year}{2006}).

\bibitem[{\citenamefont{Abate and Durian}(2006)}]{abatePRE06}
\bibinfo{author}{\bibfnamefont{A.~R.} \bibnamefont{Abate}} \bibnamefont{and}
  \bibinfo{author}{\bibfnamefont{D.~J.} \bibnamefont{Durian}},
  \bibinfo{journal}{Phys.\ Rev.\ E} \textbf{\bibinfo{volume}{74}},
  \bibinfo{eid}{031308} (\bibinfo{year}{2006}).

\bibitem[{\citenamefont{Reis et~al.}(2007{\natexlab{a}})\citenamefont{Reis,
  Ingale, and Shattuck}}]{reisPRL07}
\bibinfo{author}{\bibfnamefont{P.~M.} \bibnamefont{Reis}},
  \bibinfo{author}{\bibfnamefont{R.~A.} \bibnamefont{Ingale}},
  \bibnamefont{and} \bibinfo{author}{\bibfnamefont{M.~D.}
  \bibnamefont{Shattuck}}, \bibinfo{journal}{Phys.\ Rev.\ Lett.}
  \textbf{\bibinfo{volume}{98}}, \bibinfo{eid}{188301}
  (\bibinfo{year}{2007}{\natexlab{a}}).

\bibitem[{\citenamefont{Reis et~al.}(2007{\natexlab{b}})\citenamefont{Reis,
  Ingale, and Shattuck}}]{reisPRE07}
\bibinfo{author}{\bibfnamefont{P.~M.} \bibnamefont{Reis}},
  \bibinfo{author}{\bibfnamefont{R.~A.} \bibnamefont{Ingale}},
  \bibnamefont{and} \bibinfo{author}{\bibfnamefont{M.~D.}
  \bibnamefont{Shattuck}}, \bibinfo{journal}{Phys.\ Rev.\ E}
  \textbf{\bibinfo{volume}{75}}, \bibinfo{eid}{051311}
  (\bibinfo{year}{2007}{\natexlab{b}}).

\bibitem[{\citenamefont{Keys et~al.}(2007)\citenamefont{Keys, Abate, Glotzer,
  and Durian}}]{keysNP07}
\bibinfo{author}{\bibfnamefont{A.~S.} \bibnamefont{Keys}},
  \bibinfo{author}{\bibfnamefont{A.~R.} \bibnamefont{Abate}},
  \bibinfo{author}{\bibfnamefont{S.~C.} \bibnamefont{Glotzer}},
  \bibnamefont{and} \bibinfo{author}{\bibfnamefont{D.~J.}
  \bibnamefont{Durian}}, \bibinfo{journal}{Nature Physics}
  \textbf{\bibinfo{volume}{3}}, \bibinfo{pages}{260} (\bibinfo{year}{2007}).

\bibitem[{\citenamefont{Landau and Lifshitz}(1959)}]{lanlif7}
\bibinfo{author}{\bibfnamefont{L.~D.} \bibnamefont{Landau}} \bibnamefont{and}
  \bibinfo{author}{\bibfnamefont{E.~M.} \bibnamefont{Lifshitz}},
  \emph{\bibinfo{title}{Theory of Elasticity}} (\bibinfo{publisher}{Pergamon
  Press}, \bibinfo{address}{London, U.K.}, \bibinfo{year}{1959}).

\end{thebibliography}
\end{document}